\documentclass[journal=jacsat,manuscript=article]{achemso}

\usepackage{chemformula} 
\usepackage{tablefootnote}
\usepackage[T1]{fontenc} 

\author{Parul R. Raghuvanshi}
\author{Dipanwita Bhattacharjee}
\author{Amrita Bhattacharya}
\email{b_amrita@iitb.ac.in}
\phone{+91-22-2576-7620}
\affiliation[Indian Institute of Technology, Bombay]
{Department of Metallurgical Engineering and Materials Science, Indian Institute of Technology, Bombay, Powai-400076, Mumbai, Maharashtra, India}

\title[An \textsf{achemso} demo]
  {Self Doping for Synergistically Tuning the Electronic and Thermal Transport Coefficients in n-type Half-Heuslers}


\begin{document}


\begin{abstract}
Ternary intermetallic half-Heusler (HH) compounds (XYZ) with 18 valence electron count viz. ZrCoSb, ZrNiSn, and ZrPdSn, have revealed promising thermoelectric properties. Exemplarily, it has been experimentally observed that a slight change in the content of Y-site atoms (by $\sim$3-12.5 \% i.e., $m$ =0.03, 0.125 in ZrY$_{1+m}$Z) leads to drastic lowering in the lattice thermal conductivity $\kappa_\mathrm{L}$ by more than~65-80 \% in many of these compounds. The present work aims at exploring the possibility of maximizing the electronic transport scenario after achieving the low $\kappa_\mathrm{L}$ limit in these compounds. By taking into account the full anharmonicity of the lattice dynamics, Boltzmann transport calculations are performed under the framework of density functional theory. Our results show that these excess atoms present in the vacant lattice site induce scattering by acting either as a rattling mode or by hybridizing with the acoustic modes of the host depending upon their mass and bonding chemistry, respectively. Furthermore, the introduction of these scattering centers may lead to the formation of a defect mid-gap state in the electronic band structure (detrimental for electronic transport) or lead to light doping of the host compound. The latter is found to be particularly conducive for attaining synergy in both thermal as well as electronic transport.
\end{abstract}

\section{Introduction}
Global concerns about energy security, energy affordability and greenhouse gas emissions have heightened the interest in potential, reliable and efficient energy conversion technologies such as thermoelectrics material and/or device. A thermoelectric device typically contains pairs of $n$- and $p$-type semiconductors, generating an electric voltage from a difference in temperature $T$.~\cite{DiSalvo1999,Tritt1999,Bell2008} The energy conversion efficiency of thermoelectric material is represented by the dimensionless figure of merit, $zT= \frac{S^{2}\sigma T}{\kappa}$.~\cite{Goldsmid1964,QZhang2016} Where, $zT$ directly depends upon the power factor, PF ($S^{2}\sigma$), which is the product of square of Seebeck coefficient ($S$) and electrical conductivity ($\sigma$). On the other hand, it is inversely proportional to the thermal conductivity ($\kappa$), which comprises of the summation of electronic thermal conductivity ($\kappa_{e}$) and lattice thermal conductivity ($\kappa_\mathrm{L}$). The electronic part $\kappa_{e}$, is proportional to $\sigma$ through Wiedmann-Franz law ($\kappa_{e} = L\sigma T$) and is often smaller than the lattice part, $\kappa_\mathrm{L}$ especially for semiconductors. Naturally, for maximizing the $zT$, the synergistic enhancement of PF and reduction of $\kappa$ is the ultimate goal. However, the optimization and development of novel and improved TEMs in the laboratory is a tedious, lengthy and costly task that involves expensive prototyping and extended test series. Naturally, {\it ab-initio} calculations lend themselves to aid and guide this process: On the one hand, such simulations allow fundamental insights in the electronic and atomistic processes underlying the thermoelectric effect; on the other hand, they also allow a rapid and cost-efficient pre-screening across the whole periodic system. However, the overwhelming majority of theoretical and modelling efforts in this field have hitherto only focused on the part of the problem, i.e., either on the structural stability, or on the electronic or on the phononic transport, and thus never focussed on the crucial interplay between all these aspects. One useful strategy to enhance the efficiency in laboratory is to introduce long-range disorder via nano-structuring; however, such an approach can obviously only be successful if the basic material is already an excellent thermoelectric material (TEM) itself. Thus, the recognition of bulk and complex thermoelectric materials with intrinsically low thermal conductivity such as Clathrates,~\cite{Amrita2019,Amrita2020} Skutterudites,~\cite{Nolas1999} half-Heusler (HH) alloys,~\cite{Snyder2008, Graf2011} rare earth Chalcogenides,~\cite{Biplab2021} etc comprise a field in itself. Among these classes of materials, the 8 and 18 valence electron count (VEC) half-Heusler compounds are found to be stable ones for mid to high temperature range thermoelectric application.~\cite{Aliev1989,Aliev1990,Ogut1995,Pierre1997,Larson1999,Ciftci2016,HZhu2019} Along with their earth abundant compositions, they offer huge compositional space for reducing their $\kappa_\mathrm{L}$ (which constitutes $\sim$90 \% of their total thermal conductivity for these compounds~\cite{Henry2008, Tian2011}).~\cite{CFu2015, Holuj2015, TJZhu2017} Various approaches have been adopted for the simultaneous reduction of $\kappa_\mathrm{L}$ and improvement of $S^{2}\sigma$,~\cite{HXie2013, TJZhu2017} such as isovalent/aliovalent substitution,~\cite{Anand2019, Parul2020} co-doping at different site,~\cite{Hazama2011, Nagendra2020} introduction of a nanoscale microstructure,~\cite{Makongo2011, Kirievsky2013, Romaka2013, Birkel2013} etc. Primarily it has been observed that introduction of several different atomic species in the same unit cell (maintaining the VEC) leads to lowering in the lattice thermal conductivity.~\cite{Anand2019} However, it has been also observed that off-stoichiometric excess elemental doping of these half-Heusler phases also leads to a drastic lowering in the $\kappa_\mathrm{L}$.~\textbf{\cite{HHXie2012, Chai2015, Chauhan2019, Chauhan2019_1}} Douglas $et$ $al.$ have studied phase evolution in TiNiSn materials by adding extra Ni ($\leq$ 0.25), leading to the formation of two coexisting phases (half-Heusler-TiNiSn and full Heusler-TiNi$_2$Sn) and observed an increase in the value of $zT$ from 0.35 to 0.44~\cite{Douglas2014} at 800K. Similarly, Chai and Kimura also investigated the microstructural evolution of TiNiSn and obtained a large density of nano-sized full-Heusler (FH) precipitates within the half-Heusler matrix in slightly Ni-rich TiNiSn and found an increase in the power factor at high temperatures ($\sim$ 3.5 mWm$^{-1}$K$^{-2}$ compared to $\sim$ 2.5 mWm$^{-1}$K$^{-2}$ at 700K) for the parent compound TiNiSn.~\cite{Chai2012} The same group also investigated the effect of nano-precipitates in the HH ZrNi$_{1.1}$Sn compound and obtained a significant improvement in $zT$ (0.54 to 0.75, i.e., $\sim$40 \%) at 900 K.~\cite{Chai2015} Li $et$ $al.$ showed the phase boundary mapping, which provides an important instruction for optimizing the solubility range of interstitial Ni in the ZrNi$_{1+x}$Sn half-Heusler phase.~\cite{Li2020} These off-stoichiometric Ni-rich ZrNi$_{1+x}$Sn alloys are used to optimize for electrical and thermal transport via energy filtering, carrier concentration tuning etc. Nagendra $et$ $al.$ achieved $zT$ $\approx$ 1.1 at 873 K for the optimized ZrNi$_{1.03}$Sn composition.~\cite{Chauhan2019}. Our first-principles theoretical investigation of the experimental results also showed that self doping of Ni (by 3 \%) in ZrNiSn also leads to an unprecedented lowering in the $\kappa_\mathrm{L}$.~\cite{Nagendra2018} On the other hand, the same group examined the implication of intrinsic doping and spinodal decomposition on the thermal and electrical transport parameters of nonstoichiometric (Ti, Zr)CoSb HH systems. Where, they showed the tunning of electrical transport due to intrinsic doping as interstitial Co and defects, which are induced by excess Co off-stoichiometry in the alloys.~\cite{Chauhan2019_1}

The connection between atomic switching disorder and thermoelectric properties in ZrNiSn is investigated by Xie $et$ $al.$, along with the effects of annealing cycles on disorder present in the system and their transport parameters (both electronic and vibrational).~\cite{HHXie2012} Moreover, the formation of “in-gap” states close to the Fermi energy due to these atomic disorder is elucidated by Miyamoto $et$ $al.$.~\cite{Koji2008} and the formation of clustering and nanostructures by these excess Ni in ZrNi$_{1+x}$Sn alloys (0 $\leq$ $x$ $\leq$ 1) is thoroughly studied by D. T. $et$ $al.$~\cite{Do2014} To understand the chemical bonding motifs governing physical properties of these stoichiometric or off-stoichiometric HH compounds, Tolborg and Iversen  gave an in depth analysis of the chemical bonding, which is useful for the design of complex materials.~\cite{Tolborg2021}

The purpose of this present work is to explore the effect of excess doping on the electronic as well as thermal transport properties in the HH lattice (i.e., in one of the vacant FH sites). The earth-abundant 18 VEC HH compounds viz. ZrNiSn, ZrPdSn and ZrCoSb are chosen as prototypical exemplary cases so that the effect of both the variation in mass as well as bonding chemistry is explored in their electronic as well as vibrational structure. For this purpose, the complete compositional space starting from HH to FH limit viz. XY$_{1+m}$Z ($m$ = 0, 0.03, 0.125, 0.5, 1, and X = Zr; Y = Ni, Pd, Co; Z = Sn, Sb) is explored and their lattice thermal conductivity is considered. Once the lower limit of lattice thermal conductivity is theoretically determined, the electronic transport scenario around the given doping limit is explored to investigate the pathways to obtain maximum efficiency.

\section{Computational Details}
The density functional theory~(DFT)~\cite{Hohenberg1964, Kohn1965} calculations are performed using the Vienna Ab-initio Simulation Program (VASP),~\cite{Kresse1994, Hafner1997} which is a plane wave-based electronic structure code. For the structural optimization, the exchange-correlation potentials are described within the Generalized Gradient Approximation (GGA) with Perdew-Burke-Ernzerhof (PBE)~\cite{PBE1996} scheme. Projector Augmented Wave (PAW) method is employed and plane-wave cutoff energy of 550 eV with an energy convergence criterion of {\bf 10$^{-4}$ eV} is used. For static calculations, a converged Monkhorst-Pack k-mesh grid is applied for the unit cell. For each structure, ionic as well as geometric relaxations are performed.
In the first step, in order to examine the stability of the compositions, the formation energy ($E_\mathrm{f}$) of the compounds is calculated as;
\begin{equation}
\label{Eform}
E_\mathrm{f} = E(XY_{n}Z)- \frac{E(X_{a})}{a} -n \frac{E(Y_{b})}{b}- \frac{E(Z_{c})}{c}
\end{equation}

where, $E(XY_{n}Z)$ is the total energy (per formula unit) of the XY$_{n}$Z compound, which contains $n=1+m$ defect concentration at Y-site, and $E(X_{a})$, $E(Y_{b})$, and $E(Z_{c})$ are the total energy of the bulk phases of X, Y, and Z with $a$, $b$ and $c$ numbers of atoms respectively. Then, the effect of temperature on formation energy [$E_\mathrm{f}(T)$] is incorporated by adding the contributions stemming from the vibrational free energy $F_\mathrm{vib}$;
\begin{equation}
\label{Ef}
E_\mathrm{f}(T) = E_\mathrm{f} + F_\mathrm{vib}(T)
\end{equation}

The second term, $F_\mathrm{vib}(T)$, is the free-energy contribution of the individual compositions appearing in eqn.~\ref{Eform},
\begin{equation}
\label{Evib}
F_\mathrm{vib}(T) = F(XY_{n}Z)- \frac{F(X_{a})}{a} -n \frac{F(Y_{b})}{b}- \frac{F(Z_{c})}{c}
\end{equation}

The electrical transport characteristics, i.e., the Seebeck coefficient ($S$) and electronic conductivity ($\sigma$) are investigated from the semi-classical Boltzmann theory~\cite{Ziman1960} using the BoltzTraP package,~\cite{MADSEN200667} where scattering is treated under the constant relaxation time approximation. Fully optimized crystal structures are considered for the transport calculations, whereby a converged dense Monkhorst-Pack K-mesh is used for obtaining the energy eigenvalues.

The phonon dispersions are calculated using the finite displacement method as implemented in the phonopy,~\cite{Togo2015} which is a post-processing python-based code for calculating the vibrational properties of solids. Converged supercells are used for the respective unit cells to calculate the phonon band structures. The amplitude of the displacements is fixed to 0.01 \AA, and the forces are converged to the accuracy of 10$^{-8}$eV/\AA. The phonon group velocity $v$ ($v= \mathrm{d\omega}/\mathrm{dK}$), mode resolved phonon group velocity ${v_i}$~($v_i= \mathrm{d\omega}_i/\mathrm{dK}$), Gr$\ddot{\mathrm{u}}$neisen parameter $\gamma$~($\gamma = -\frac{\mathrm{V_0}}{\omega}\frac{\mathrm{d\omega}}{\mathrm{dV}}$), and mode resolved Gr$\ddot{\mathrm{u}}$neisen parameter $\gamma_i$ ($\gamma_i= -\frac{\mathrm{V_0}}{\omega_i}\frac{\mathrm{d\omega_i}}{\mathrm{dV}}$) are extracted from the harmonic phonon band dispersion using self-written python-based extensions to phonopy-VASP. For calculating the $\gamma$ and $\gamma_i$, the lattice is subjected to strains of $\pm$ 2~\%. Finally, the Asen-Palmer modified version of the Debye Callaway theory~\cite{Callaway1959} parameterized for solids~\cite{Asen1997,YZhang2012,Nagendra2018,Amrita2018} is used to calculate $\kappa_\mathrm{L}$ of the compounds from their harmonic phonon dispersion. Simultaneously, Boltzmann transport equation (BTE) for phonons under relaxation time approximation (RTA) is solved using ShengBTE package,~\cite{ShengBTE_2014} whereby the 2$^{\mathrm{nd}}$ and 3$^{\mathrm{rd}}$ order interatomic force constants (IFCs) is evaluated to calculate the $\kappa_\mathrm{L}$. Finally, the effect of anharmonicity arising from the three phonon scattering processes is also incorporated at a given temperature. 

\section{Results and discussion}
\subsection{Structure and stability}
\label{sec:SS}
\begin{figure}
\centering
\includegraphics[scale=0.5]{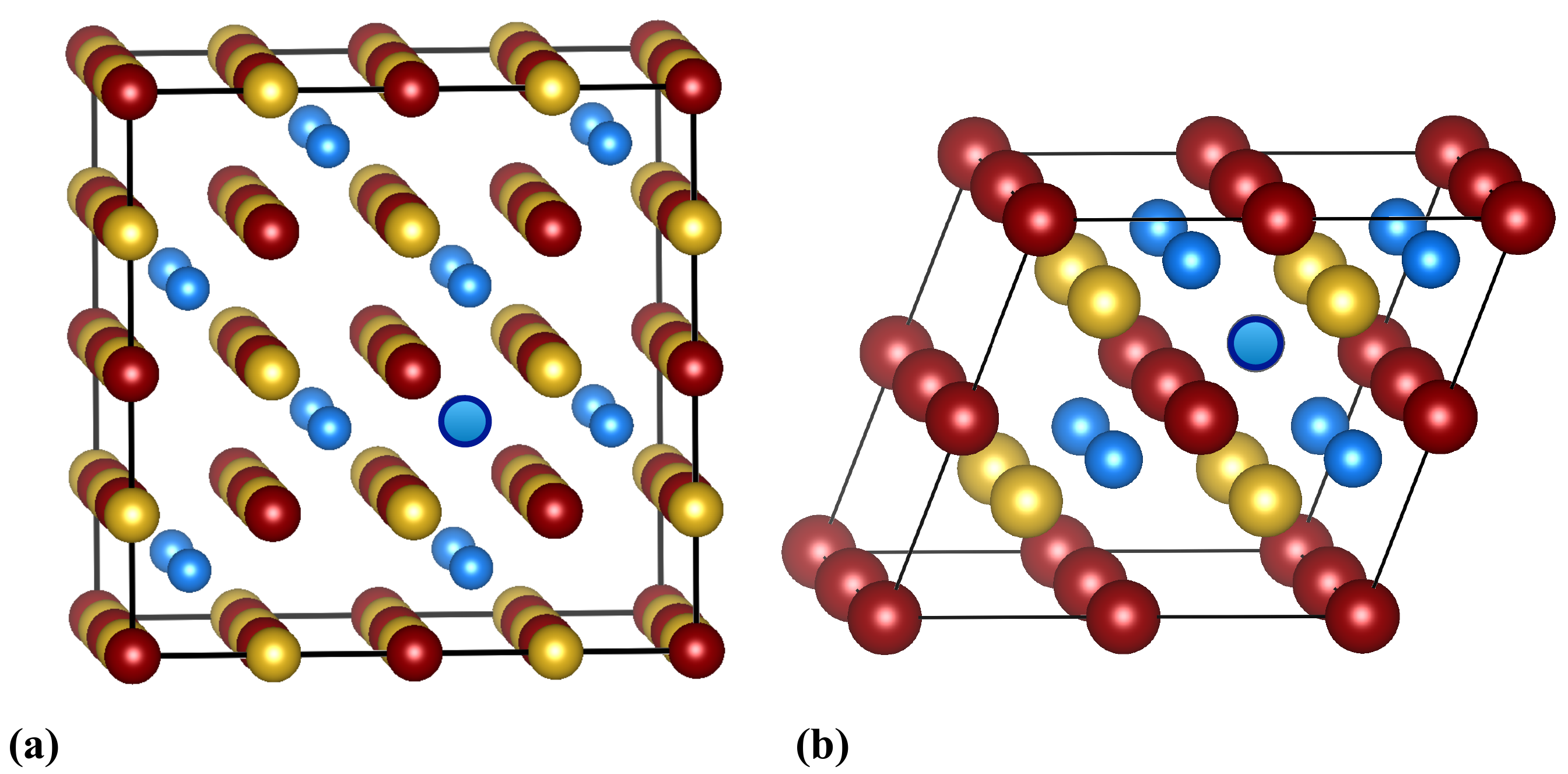}
\caption{The supercell used for ZrY$_{1+m}$Z i.e. for (a) $m$ = 0.03 (generated from the 2 $\times$ 2 $\times$ 2 conventional cubic structure) (b) $m$ = 0.125 (generated from 2 $\times$ 2 $\times$ 2 primitive cell). The Zr, Y and Z atoms are shown in maroon, blue and yellow colors, while the outlined blue circle denotes a tentative full Heusler lattice site, which can accommodate the extra atom.}
\label{fig:structure}
\end{figure}

\begin{figure}
\centering
\includegraphics[scale=0.78]{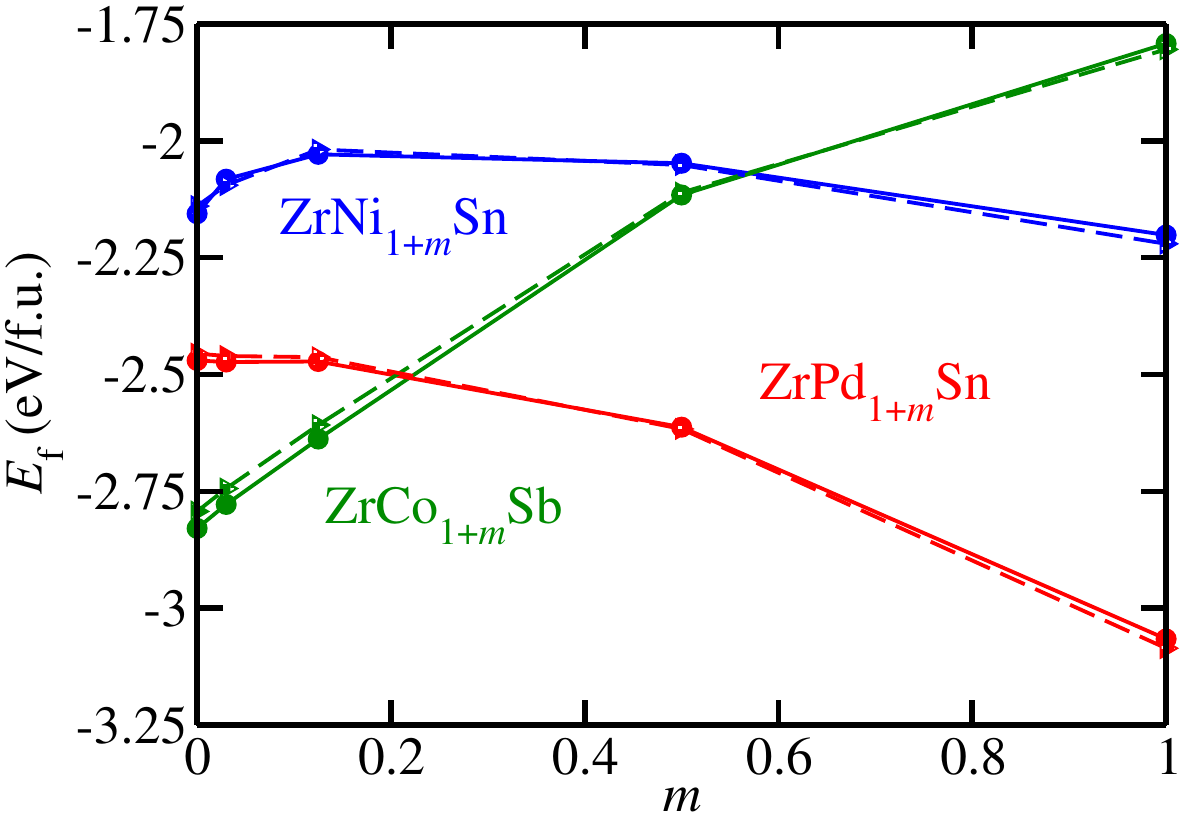}
\caption{Formation energy $E_\mathrm{f}$ of ZrY$_{1+m}$Z (Y = Ni, Pd, Co; Z = Sn, Sb) plotted as a function of defect concentration, $m$ (viz. $m$ = 0, 0.03, 0.125, 0.5, 1). The effect of the zero point energy is also incorporated (dashed line) and compared with the static energy calculation (solid line).}
\label{fig:FE}
\end{figure}

Heusler alloys (i.e., HH, with one vacant site per formula unit, XYZ and FH, with all sites occupied, XY$_2$Z) are intermetallic compounds with four unique high symmetry positions in a unit cell. They have four different Wyckoff positions viz. $4a$ (0, 0, 0), $4c$ (1/4, 1/ 4, 1/4), $4d$ (3/4, 3/ 4, 3/4) and $4b$ (1/2, 1/2, 1/2) for accommodating the atoms.~\cite{Graf2011, Zeier2016} The compositions are chosen to satisfy the 18 total VEC rule viz. the half-Heusler ZrNiSn, its isovalent variant ZrPdSn (to explore the mass variation) and ZrCoSb (in order to explore the variation in bonding chemistry). Starting from the HH (XYZ) composition, the concentration of the element at the Y site is varied until its FH (XY$_{2}$Z) limit (i.e., XY$_{1+m}$Z, where $m$ = 0, 0.03, 0.125, 0.5, 1). In order to generate the off stoichiometric compositions viz. 0.03, 0.125, 0.5, supercells of conventional and primitive structures containing the different number of atoms in the lattice are used [see Fig.~S2 for $m$ = 0 in SI]. An iterative scanning of the high symmetry position is employed to explore the lowest energy composition. For example, a composition XY$_{1.03}$Z is obtained by adding one extra Y atom to one of the $4d$ sites of 2 $\times$ 2 $\times$ 2 supercell of the conventional unit cell of HH (containing 96 atoms: 32 X, 32 Y, and 32 Z), as shown by outlined blue circle in Fig.~\ref{fig:structure} (a). Using this iterative strategy for filling the subsequent low energy sites, the XY$_{1.5}$Z composition is also obtained. On the other hand, for generating the XY$_{1.125}$Z composition, one extra Y atom is added to the 2 $\times$ 2 $\times$ 2 supercell (containing 24 atoms; 8 X, 8 Y, and 8 Z) of the primitive cell of the HH compound \textbf{[see Fig.~\ref{fig:structure} (b)]}. An analysis of change in lattice parameter and space group with doping is given in the supplementary (see SI section S2, Table S1), the lattice parameter is typically found to be increased with doping concentration, which is found to be in agreement with experimental literature.~\cite{Do2014, Page2015, Chauhan2019}

The stability of these compositions i.e., ZrY$_{1+m}$Z (Y = Ni, Pd, Co; Z = Sn, Sb and $m$ = 0, 0.03, 0.125, 0.5, 1) is analyzed by calculating the formation energies using eqn~\ref{Eform}. The thermodynamic phase stability at 0 K (see Fig.~\ref{fig:FE}) is calculated by including the zero point vibrational energy of the compositions, whereby the $E_\mathrm{f}(T)$ is plotted as a function of doping concentration [see section S1 in SI for $E_\mathrm{f}(T)$ plotted as a function of $T$]. In the case of ZrNi$_{1+m}$Sn, the end members are found to be slightly more stable than the off stoichiometric compositions, as indicated by their more negative formation energy, which is in agreement with the previous theoretical study~\cite{Romaka2013}. For ZrPdSn, the increase in Pd concentration (in ZrPd$_{1+m}$Sn) lead to stabilization of the phases. But only in the case of ZrCo$_{1+m}$Sb, the relative stability of the phases is found to be lowered with an increase in doping concentration. However, all the stoichiometric as well as the off-stoichiometric compositions are found to have negative formation energy, which indicates their formation in the laboratory through some non-equilibrium routes. Subsequently, the effect of doping on the electronic and vibrational transport properties of the compositions is analyzed.

\subsection{Vibrational spectrum and thermal transport}

\begin{figure}
\centering
\includegraphics[scale=.45]{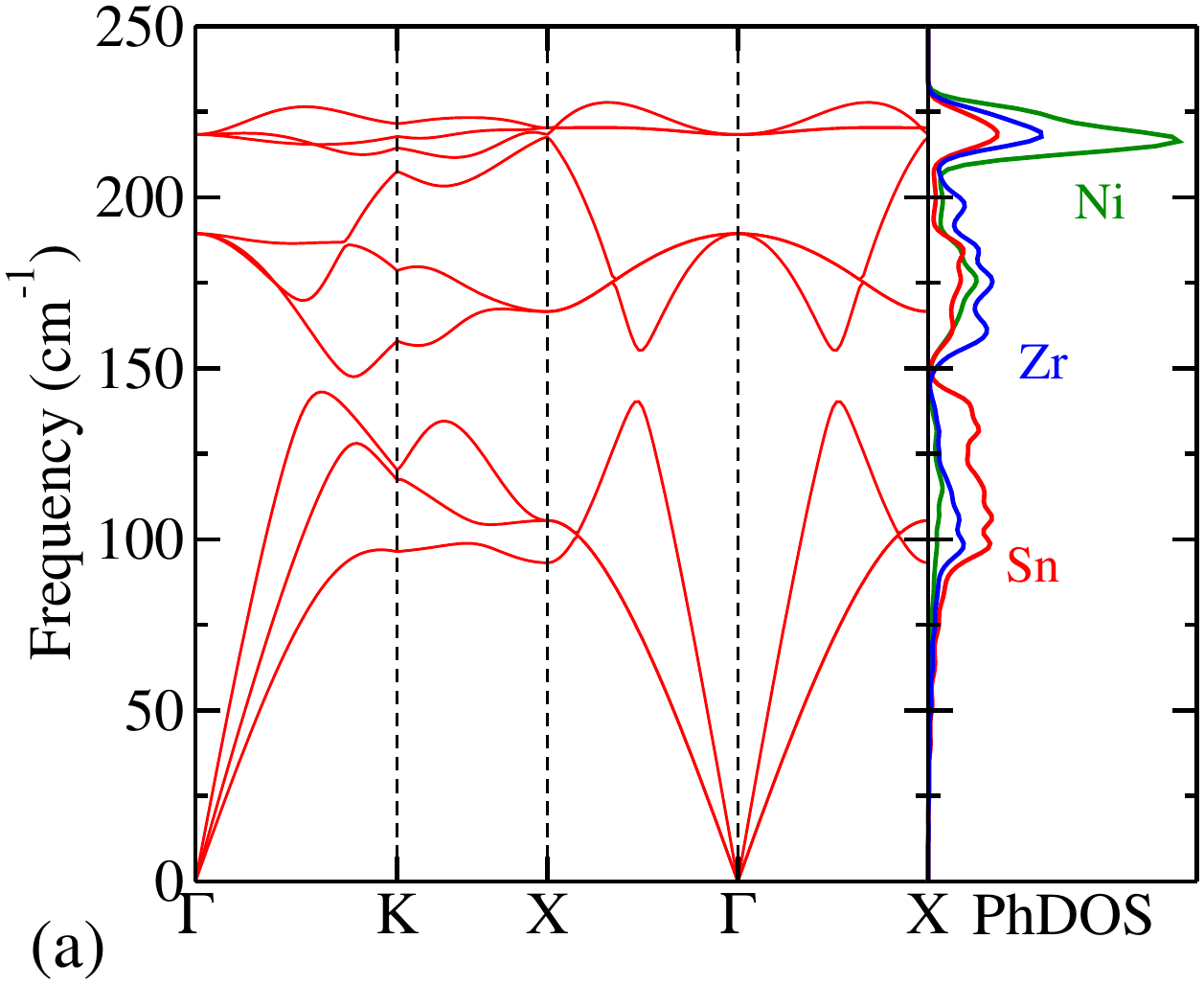}
\vspace{0.35cm}
\includegraphics[scale=.45]{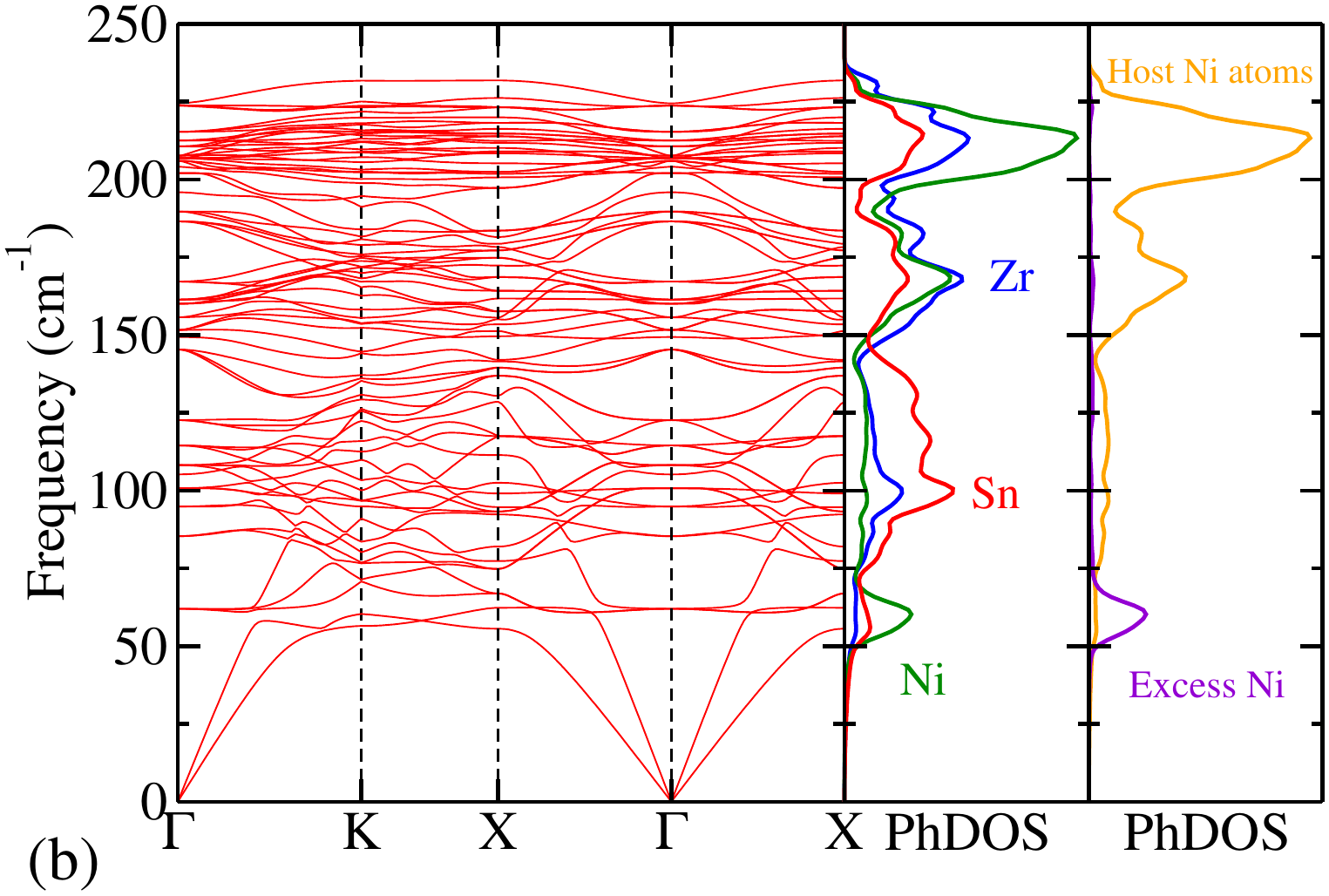}
\includegraphics[scale=.45]{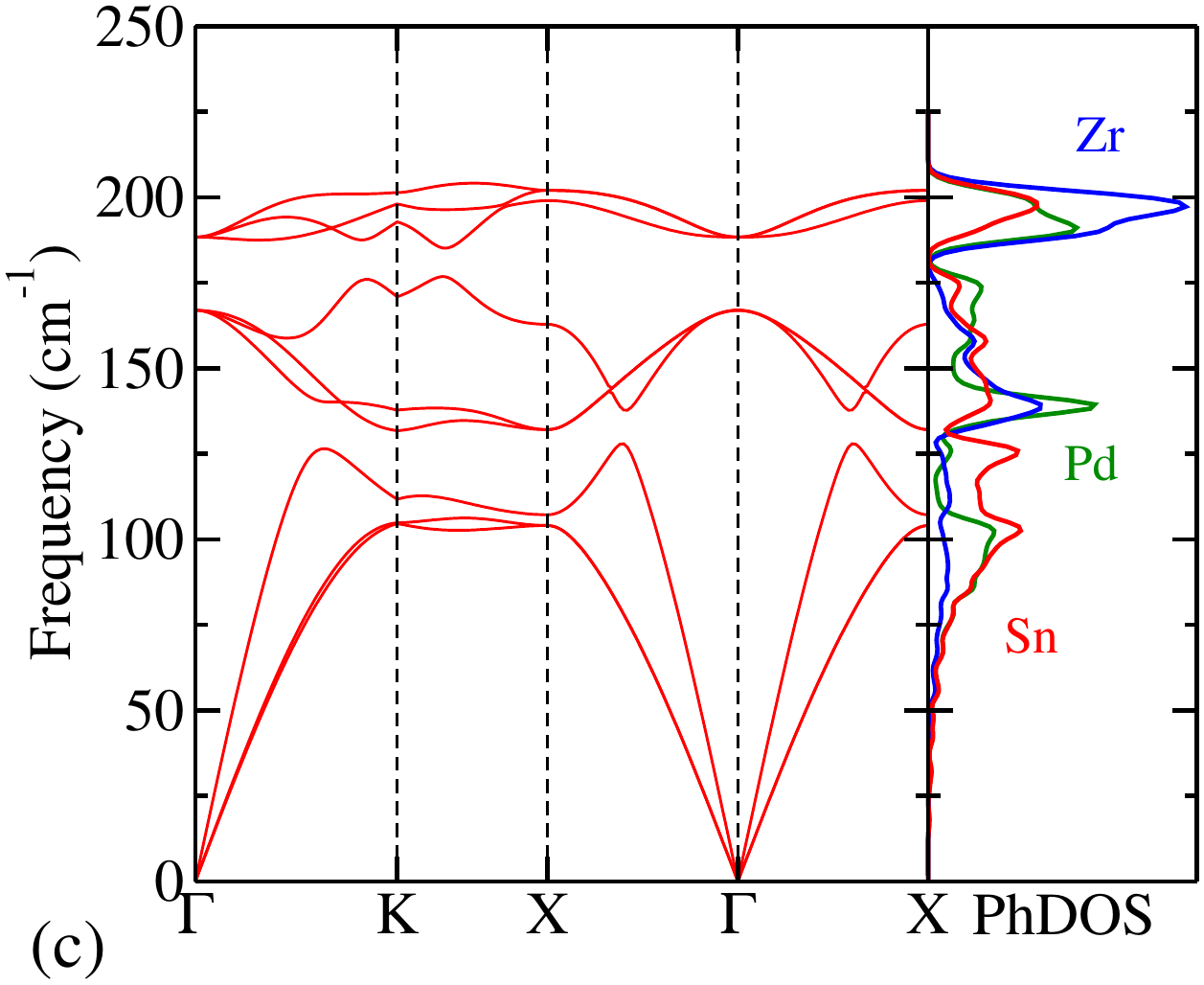}
\vspace{0.35cm}
\includegraphics[scale=.45]{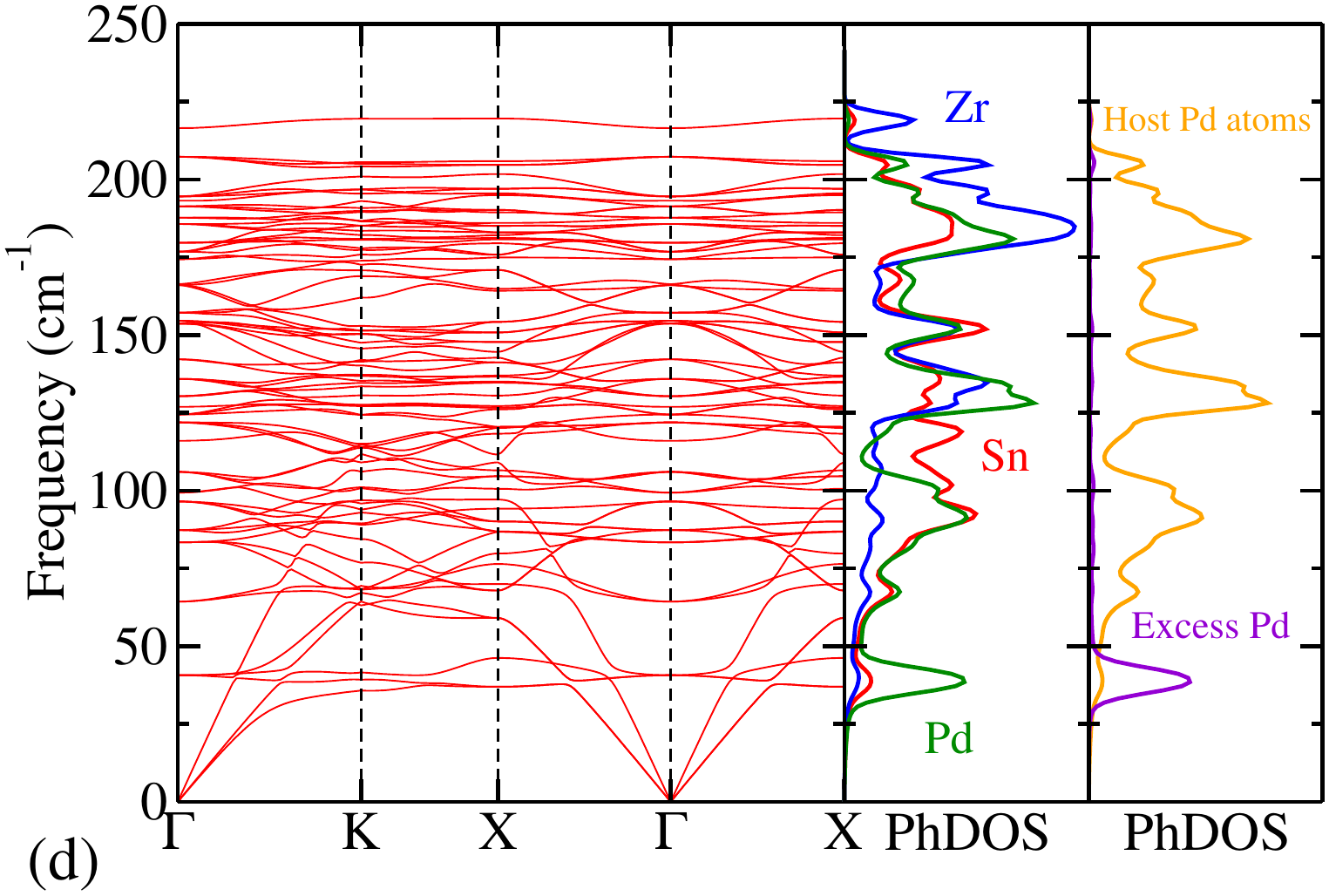}
\includegraphics[scale=.45]{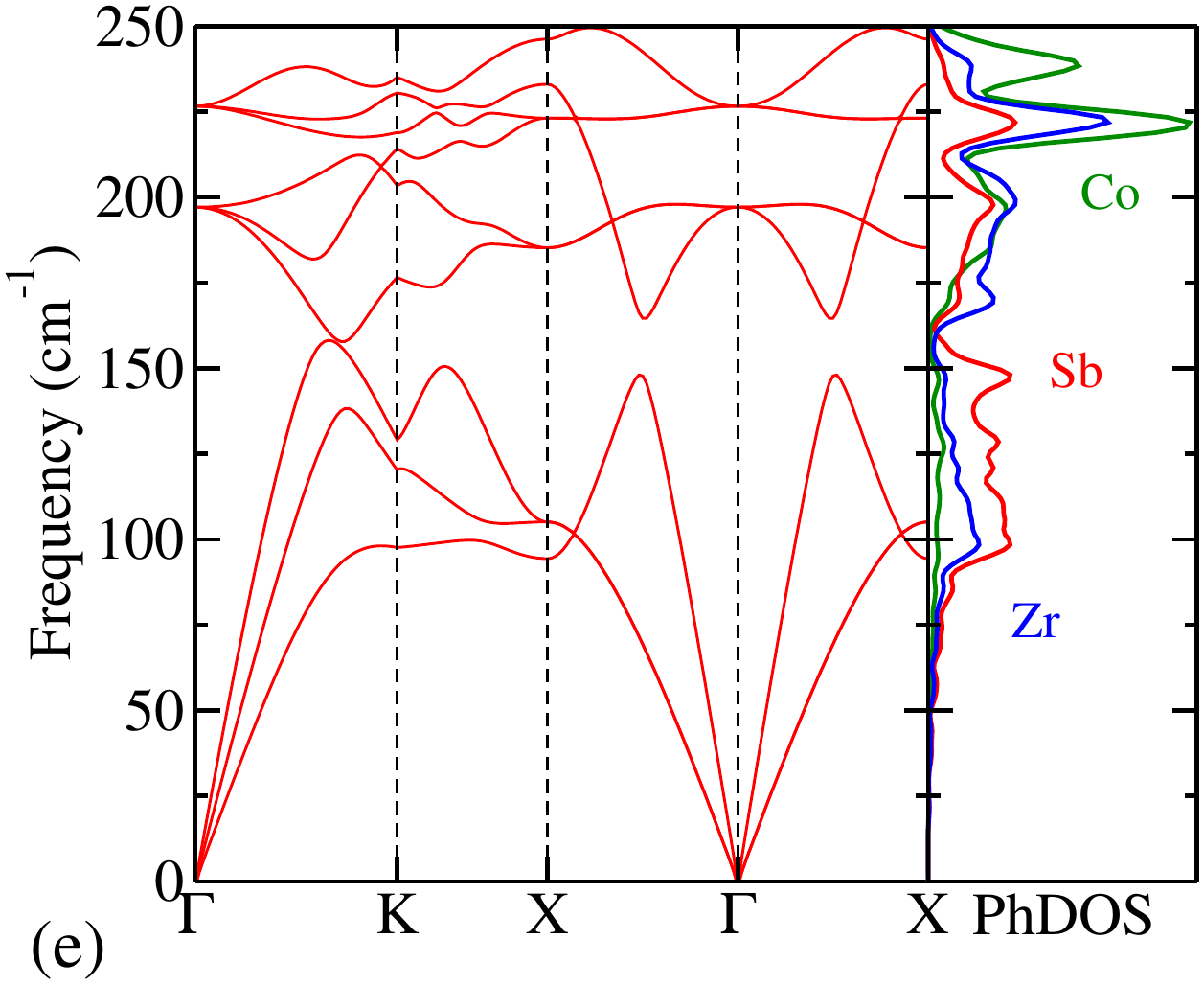}
\includegraphics[scale=.45]{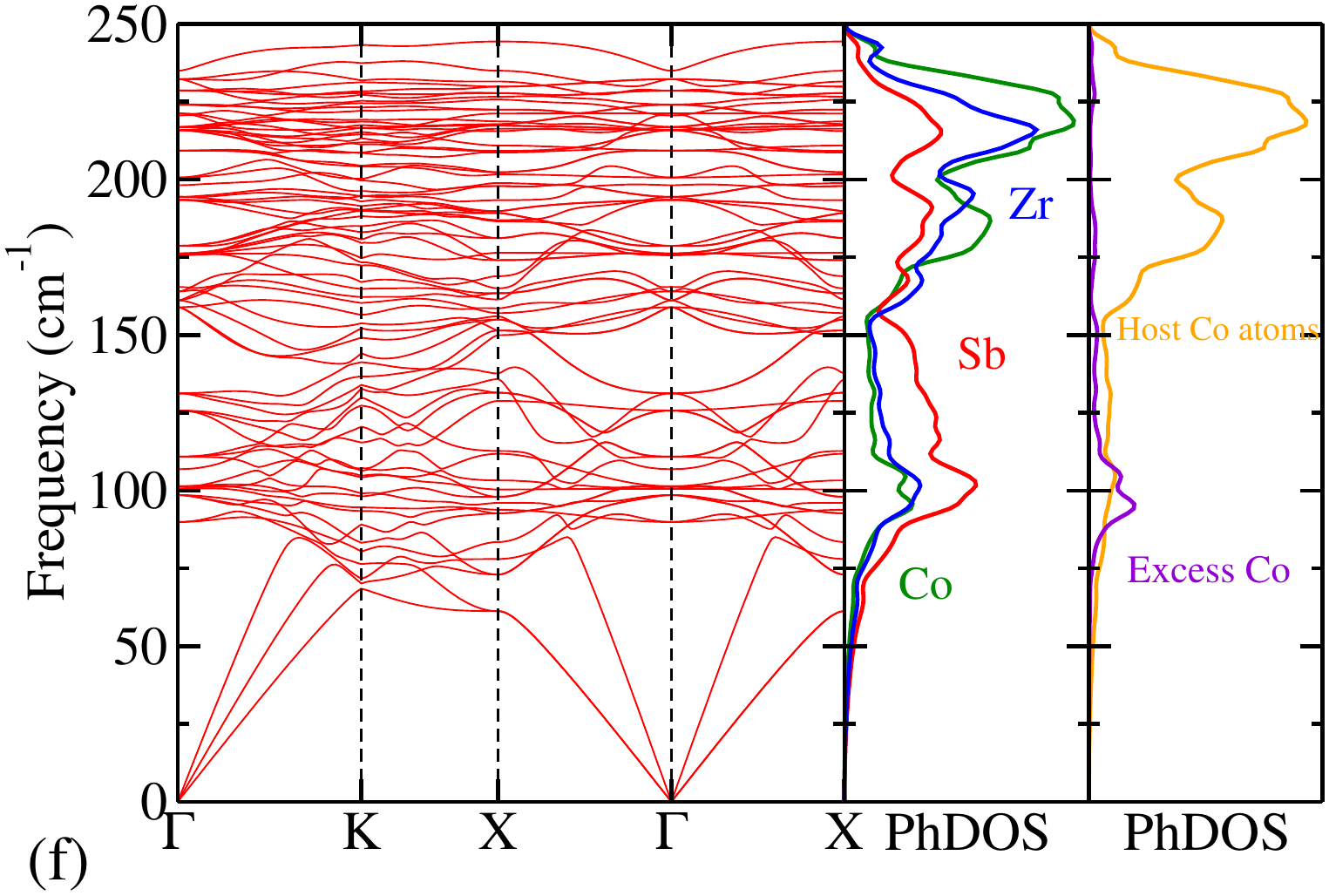}
\caption{Phonon band dispersion and partial phonon density of states of the pristine and the 12.5 \% doped composition of (top) ZrNiSn (a-b), (middle) ZrPdSn (c-d) and (bottom) ZrCoSb (e-f). In order to achieve the 12.5 \% doping, a 2 $\times$ 2 $\times$ 2 of primitive cell structure containing 8 X, 9Y and 8 Z atoms. In addition the phonon convergence is also checked by employing supercell method. For the 12.5 \% compositions, the end right panel in all cases show the partial density of states of stemming from the host Y atoms (orange) and from the extra Y atom (violet).}
\label{fig:kl}
\end{figure}

The main focus of this study is to attain the synergy between thermal transport and electronic transport. For this purpose, the vibrational transport properties is analyzed in the compositions XY$_{1+m}$Z ($m$ = 0, 0.03, 0.125, 0.5, 1, and X = Zr; Y = Ni, Pd, Co; Z = Sn, Sb). The idea is to filter the compounds with low $\kappa_\mathrm{L}$ before subsequently optimizing their electronic power factor. The phonon spectrum and phonon partial density of states (PhDOS) of the stoichiometric (HH) as well as non-stoichiometric (12.5 \%) doped compositions are shown in Fig.~\ref{fig:kl} (a-f) (see SI section S3 for the phonon spectrum of the 3 \% compositions). The highest frequency of the acoustic phonon modes extends up to $\sim$150 cm$^{-1}$ in the parent XYZ compounds [Fig.~\ref{fig:kl} (a), (c), (e)]. These acoustic modes transfer the maximum amount of energy ($\sim$90 \%) required for thermal conduction\textbf{~\cite{Henry2008, Tian2011}}. However, in all the doped compositions, the introduction of the defect atoms leads to suppression of the acoustic phonon modes, which reach the Brillouin zone boundary with much lower energy. As a result, in ZrNi$_{1.125}$Sn and ZrPd$_{1.125}$Sn compounds, the frequency at which the acoustic modes reach the zone boundary is reduced by $\sim$45 \% and $\sim$60 \% respectively. In these two cases, the suppression of the acoustic branches is brought about by the localized modes (at $\sim$60 cm$^{-1}$ and $\sim$40 cm$^{-1}$ frequency for ZrNi$_{1.125}$Sn and ZrPd$_{1.125}$Sn respectively) of the extra atom (also known as the `rattling' modes), which do not couple with the host lattice. Thus, the frequency of these localized rattling modes is found to vary as a function of the mass of the extra dopant atom i.e., $\sim$30 \% suppressed in frequency scale for heavier Pd host compared to the lighter Ni host. This, as will be seen later, acts as a tool for lowering of the $\kappa_\mathrm{L}$ due to the lowering of the Debye temperature of the individual acoustic modes (which corresponds to the maximum frequency at which the phonon mode reach the zone boundary) as enlisted in Table~\ref{tab:kl}. A similar trend is also observed for the ZrY$_{1.03}$Sn (Y = Ni, Pd) compositions [see SI in section S3, Fig.~S3 (a), (b) for details]. The lowering in acoustic modes is also observed in ZrCo$_{1.125}$Sb (by $\sim$35 \% compared to its parents), but the mechanism is found to be different from the other two prototypical cases. In this case, an isolated rattling mode of the defect is not observed in the phonon spectrum. Instead, the excess Co atoms hybridize with the acoustic as well as the low-lying optical phonon modes and spread over a long-range of frequency [see the partial PhDOS plot of Fig.~\ref{fig:kl} (f)]. In case of lower concentration of Co doping, the spread is observed from a lower frequency in the acoustic region [see SI in section S3, Fig~S3 (c)]. 

The thermal properties viz. mode-specific Debye temperatures ($\mathrm{\theta}_\mathrm{i}$), sound velocity [$v_\mathrm{s}= [\frac{1}{3}(\frac{1}{{v_\mathrm{LA}}^3} +\frac{1}{{v_\mathrm{TA}}^3} +\frac{1}{{v_\mathrm{TA'}}^3})]^\frac{-1}{3}$], Debye temperature [$\mathrm{\theta}_\mathrm{D}$ = $\frac{\hbar}{k_{B}}\left (\frac{3N}{4\pi V} \right )^{1/3}\upsilon _{s}$], and mode-specific Gr$\ddot{\mathrm{u}}$neisen parameter ($\gamma_i$) for the different compositions of the different compounds (ZrY$_{1+m}$Z with $m$ = 0, 0.03, 0.125 for Y = Ni, Pd, Co and Z = Sn, Sb) as extracted from the phonon spectrum are compared in Table~\ref{tab:kl}. On comparing the $\mathrm{\theta}_\mathrm{i}$s of parent HH with the other compositional variants i.e., ZrY$_{1+m}$Z ($m$ = 0.03, 0.125 and Y = Ni, Pd, Co; Z = Sn, Sb), a substantial variation is observed in individual $\mathrm{\theta}_\mathrm{i}$s in the doped compound, however not much difference is observed in $\mathrm{\theta}_\mathrm{D}$ values. This slight variation in $\mathrm{\theta}_\mathrm{D}$ is because of the similar group velocity (varying by $\leq$10 \%) of the compositions. The $\gamma_i$s offer an insight into the effect of anharmonicity in the vibration of the crystal i.e., a perfectly harmonic crystal should have $\gamma_i$s close to one. As expected, in all cases, the transverse (TA/TA') acoustic modes are found to have lower anharmonicity than the longitudinal (LA) acoustic modes. In the case of the parent HH compounds, ZrNiSn is found to show the lowest mode anharmonicity, while ZrPdSn is found to have the highest mode anharmonicity. However, the anharmonicity is found to vary differently with the change in doping concentration in different host HH compounds. For ZrPd$_{1+m}$Sn, the anharmonicity is found to decrease with an increase in doping concentration since the doping leads to stabilization of the lattice (see Fig.~\ref{fig:FE}). For ZrNi$_{1+m}$Sn, the doping leads to increasing anharmonicity of the modes, while for ZrCo$_{1+m}$Sb, the modal anharmonicity remains constant with doping. 

\begin{table}
\caption{\label{tab:kl}
The mode resolved~($v_i$) and average~($v_\mathrm{s}$)~phonon group velocity~(Km/s), mode resolved Gr$\ddot{\mathrm{u}}$neisen parameter~($\gamma_i$), mode resolved~($\mathrm{\theta}_{i}$) and average ($\mathrm{\theta}_\mathrm{D}$) Debye temperature~(K) of the transverse acoustic~($i$ = TA/TA$'$) and longitudinal acoustic~($i$ = LA) branches. $v_i$ and $\gamma_i$ are calculated from the maximum group velocity and Gr$\ddot{\mathrm{u}}$neisen parameter attained by the phonon modes near the $\Gamma$ point. $v_\mathrm{s}$ is calculated from $v_\mathrm{TA}$, $v_\mathrm{TA'}$, and $v_\mathrm{LA}$ using the expression $v_\mathrm{s}= [\frac{1}{3}(\frac{1}{{v_\mathrm{LA}}^3} +\frac{1}{{v_\mathrm{TA}}^3} +\frac{1}{{v_\mathrm{TA'}}^3})]^\frac{-1}{3}$. Finally, the $\kappa_\mathrm{L}$~(Wm$^{-1}$K$^{-1}$) is calculated using the Debye Callaway (DC) formalism and the corresponding value at 300 K is enlisted for few selective compounds for the different compositional viz. XYZ, XY$_{1.03}$Z, and XY$_{1.125}$Z. The $\kappa_\mathrm{L}$ calculated (at 300 K) by solving the Boltzmann transport equation using the three phonon scattering process in the harmonic as well as anharmonic regime as calculated using ShengBTE (BTE) code is also incorporated. The $\kappa_\mathrm{L}$ values as obtained from the literature for the two given methods are compared in bracket. The corresponding  experimental values (if available) is included in the last column.}
\renewcommand{\arraystretch}{1.1}
\begin{tabular}{p{2 cm}p{0.8 cm}p{0.78 cm}p{0.78 cm}p{0.8 cm}p{0.8 cm}p{0.8 cm}p{1.2cm}p{1.2cm}p{1cm}}
\hline
System& $i$ & $v_i$  & $v_\mathrm{s}$ & $\gamma_i$ & $\theta_i$ &$\theta_\mathrm{D}$  &  $\mathrm{\kappa}_\mathrm{L}\mid^{\mathrm{DC}}_{300}$ &$\mathrm{\kappa}_\mathrm{L}\mid^{\mathrm{BTE}}_{300}$ & $\mathrm{\kappa}_\mathrm{L}\mid^{\mathrm{Exp.}}_{300}$ \\     
\hline
               & TA           &2.7    &          & 1.2   & 136 \\
ZrCoSb &  TA$'$        &3.9   & 3.5    &1.7    & 159 &    388 & 10.5 & 23.0 &11.5$^{\cite{Chauhan2019_1}}$ \\
              & LA            &6.3   &           &2.5    &165&&&(25.0)$^{\cite{Anand2019}}$ \\     
\hline
                   & TA      & 2.8 &             &1.1  &  135 \\
ZrNiSn      & TA$'$   &  3.6 &   3.4   & 1.1  &  158 &   382  & 10.0 & 19.8 &6.2$^{\cite{Chai2015}}$\\
                  &LA       &  5.7 &           & 1.7 & 160&&(11.5)$^{\cite{Nagendra2018}}$&(20.0)$^{\cite{Anand2019}}$&4.5$^{\cite{Chauhan2019}}$\\
\hline                 
             & TA        & 2.7 &         & 1.3  & 150\\
ZrPdSn  & TA$'$  & 2.8 & 3.1 &  2.2 & 150 &  327 &  8.0 & 19.6 \\
             & LA        & 5.0 &        & 2.3  & 157\\
\hline
                           & TA           &3.1    &          & 1.2  & 73 \\
ZrCo$_{1.03}$Sb& TA$'$     &3.5   & 3.6     &1.3   & 78 &    407 & 3.7 & 4.6 & 11.0$^{\cite{Chauhan2019_1}}$ \\
                           & LA           &5.5   &           &1.7   & 82\\
\hline
                                                 & TA       & 3.0    &       &1.2  &  70 \\
ZrNi$_{1.03}$Sn & TA$'$ &  3.2 & 3.4 & 1.6  &  82 & 379 & 3.6 & 4.2&3.4$^{\cite{Chauhan2019}}$ \\
                                                 & LA       &  5.4.  &       & 1.0  & 52 &&(3.2)$^{\cite{Nagendra2018}}$\\
\hline                 
                            & TA       &   2.4 &      & 1.0  & 52\\
ZrPd$_{1.03}$Sn  & TA$'$  & 2.7    & 2.8 & 1.3  & 57 &  299 &  2.0 & 3.0 \\
                             & LA      & 5.0     &      & 1.7 & 58 \\
\hline
                             & TA           &2.6  &           & 1.3    & 92\\
ZrCo$_{1.125}$Sb & TA$'$     &3.5     & 3.2  &1.7     & 103&  365  & 4.3 & 3.4\\
                              & LA          &5.7    &           &2.0    &104\\
\hline
                                                    &TA        & 2.6 &         &1.4  &  79\\
ZrNi$_{1.125}$Sn & TA$'$ & 3.2 & 3.2 & 1.8  &  86 & 355 & 2.7 &1.6&4.6$^{*}$$^{\cite{Chauhan2019}}$\\
                                                     &LA       & 5.3 &         & 1.9    & 88&&&&5.0$^{*}$$^{\cite{Chai2015}}$\\
\hline                 
                              &TA        & 1.5 &        & 1.5 & 52 \\
ZrPd$_{1.125}$Sn  & TA$'$  & 1.6 & 3.5 &  1.6 & 62 &  376 &  1.4 & 1.1\\
                               & LA       & 2.0  &        & 2.0  & 64\\
\hline
\hline
\end{tabular}
\end{table}

\begin{figure}
\centering
\includegraphics[scale=0.55]{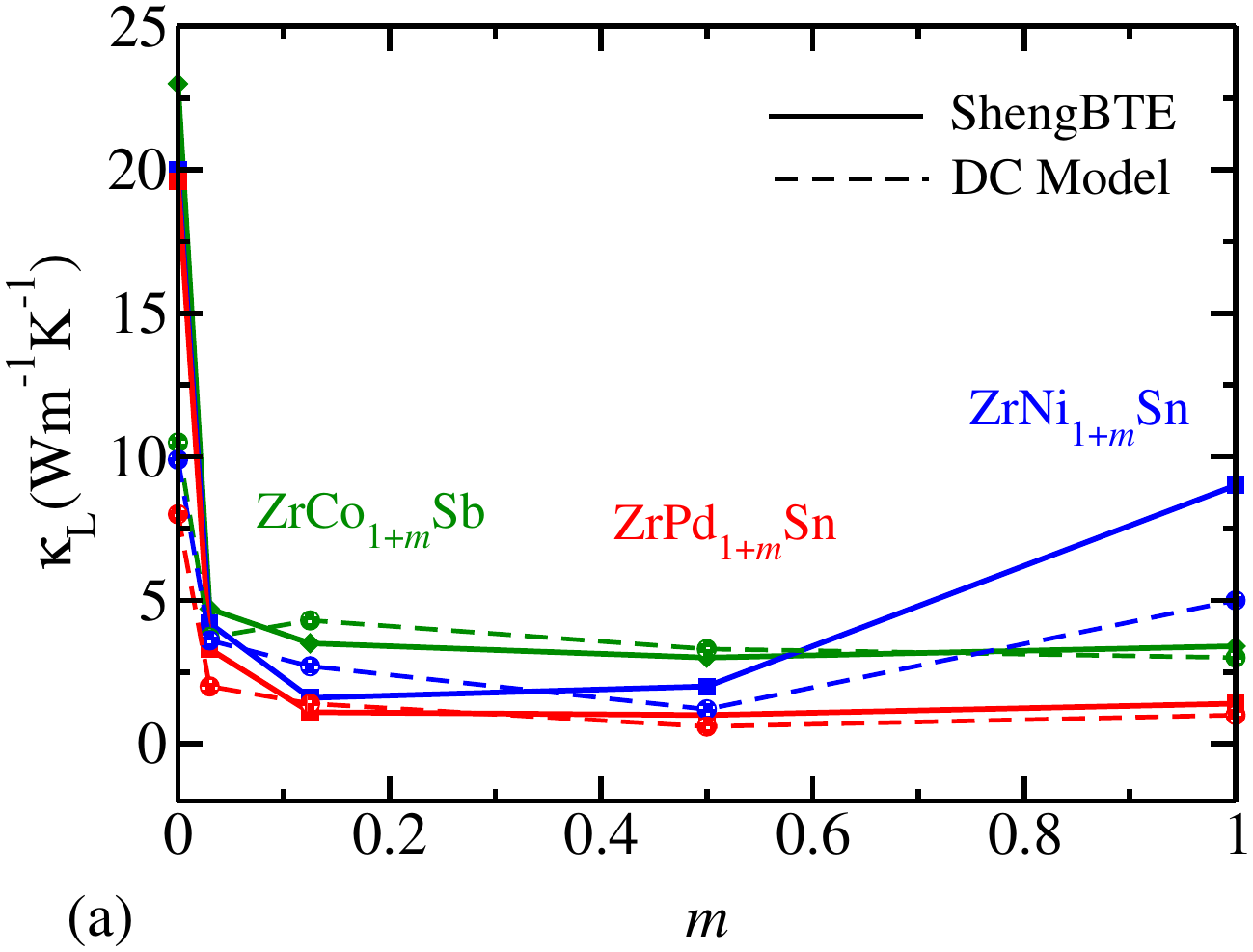}
\includegraphics[scale=0.55]{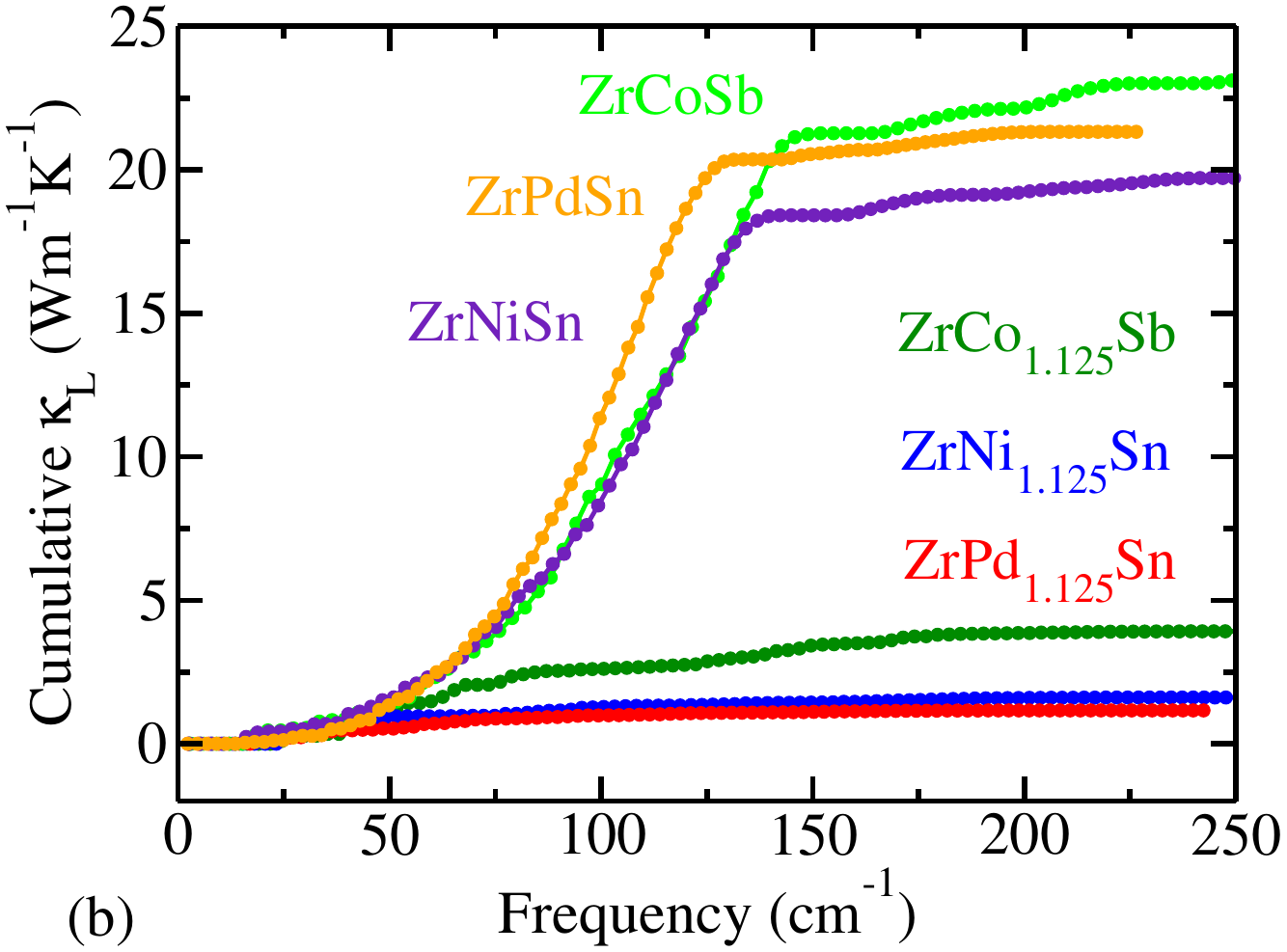}
\includegraphics[scale=0.75]{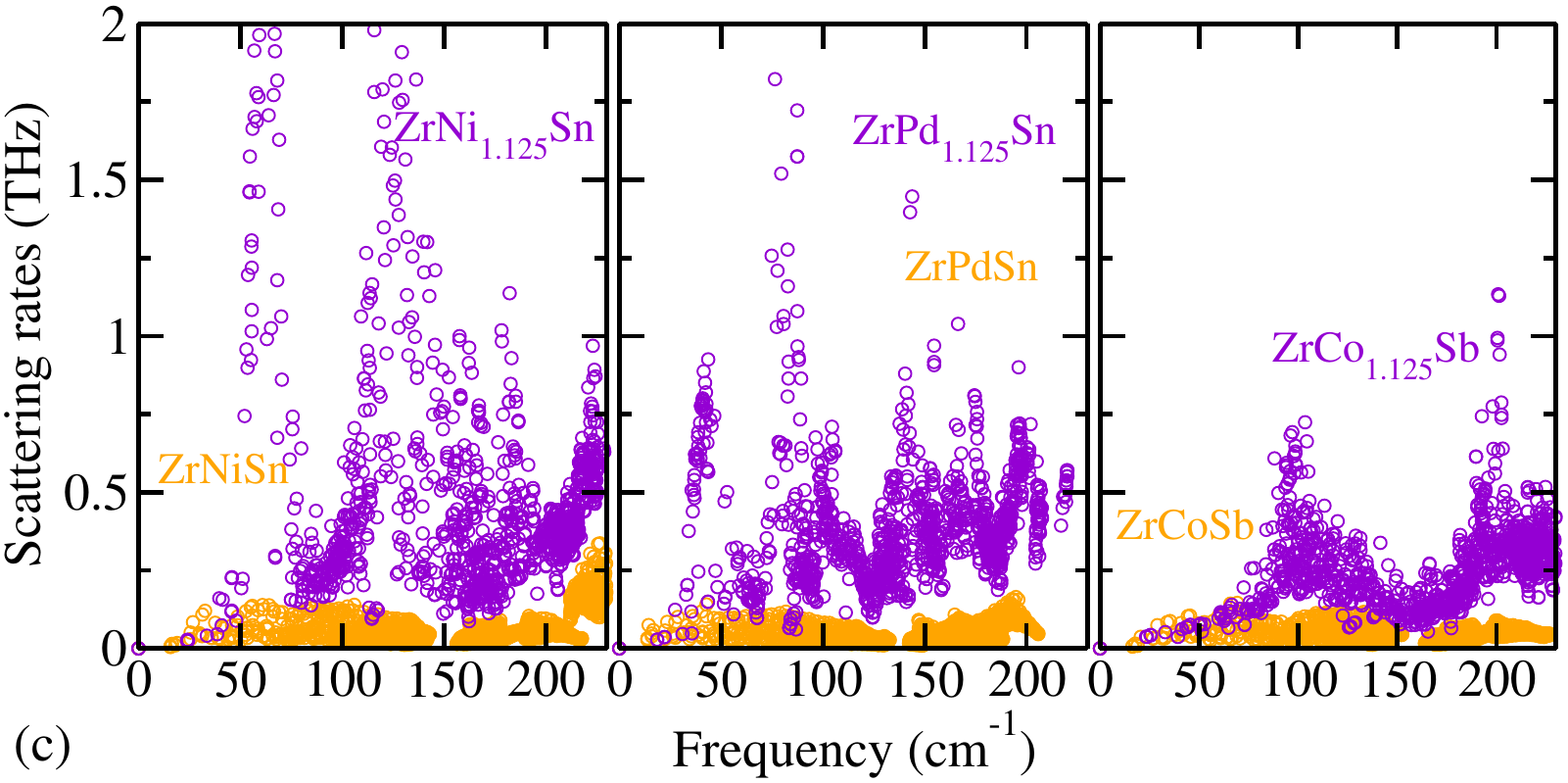}
\caption{(a) Lattice thermal conductivity ($\kappa_\mathrm{L}$) of ZrY$_{1+m}$Z (for $m$ = 0, 0.03, 0.125, 0.5, 1 and Y = Ni, Pd, Co; Z = Sn, Sb) as a function of excess defect concentration $m$. The (b) cumulative lattice thermal conductivity and (c) scattering rates in ZrY$_{1+m}$Z (for $m$ = 0, 0.125, Y = Ni, Pd, Co; Z = Sn, Sb) are also plotted as a function of phonon frequency at room temperature.}
\label{fig:rate}
\label{fig:kappa}
\end{figure}

The parameters enlisted in Table~\ref{tab:kl} are used to calculate the normal and Umklapp scattering lifetimes using the modified Debye Callaway model (see SI, section S4), which can be in turn used to calculate the $\kappa_\mathrm{L}$ for the different compositions~[see Fig.~\ref{fig:kappa} (a)]. The $\kappa_\mathrm{L}$ is found to be drastically lowered (up to $\sim$65-80\% depending upon the parents) as soon as the defect is introduced (i.e., at $m$ = 0.03). For all the compositions, it attains its lowest value at the doping concentration of $m$ = 0.125, beyond which it either remains constant (i.e., in case of ZrPd$_{1+m}$Sn and ZrCo$_{1+m}$Sb) or increase (i.e., in case of ZrNi$_{1+m}$Sn) upon further doping. The calculated values of $\kappa_\mathrm{L}$ are in good agreement with previous literature and are also in compliance with other experimental results.~\cite{Douglas2014, Nagendra2018} Given that the maximum change is observed near the doping limit of 3-12.5 \%, the phonon spectrum of one of the relevant doping concentrations (i.e.,12.5 \%) is compared with the parent HH in Fig.~\ref{fig:kl}. The scattering lifetime is directly proportional to powers of $v_\mathrm{i}$s and is inversely proportional to the square of the $\gamma_i$s (see SI, section 3). As implied from the values in the table, the group velocity do not play a significant role in dictating the trend of the $\kappa_\mathrm{L}$. On the other hand, the mode Debye temperature $\mathrm{\theta}_\mathrm{i}$s limit the expression for the $\kappa_\mathrm{L}$. Thus, the $\mathrm{\theta}_\mathrm{i}$s as well as anharmonicity (as will be discussed in detail in the following section) are found to play a pivotal role. 

In order to explore the effect of full anharmonicity in the scattering processes, the Boltzmann transport equation for phonon is solved using the 2$^{\mathrm{nd}}$ and 3$^{\mathrm{rd}}$-order interatomic force constants as implemented in the ShengBTE code. In this case, all the three phonon scattering processes (due to the normal as well as Umklapp scattering) arising due to the vibration of the lattice coupled with the defects. In the case of the off stoichiometric compositions, the dominant source of defect scattering is the extra atom, which is introduced at one of the empty 4$d$- sites of the HH structure. These scattering mechanisms inhibit the motion of phonon in the lattice, resulting in the reduction of lattice thermal conductivity. The $\kappa_\mathrm{L}$ of compositions [viz. ZrY$_{1+m}$Z (Y = Ni, Pd, Co; Z = Sn, Sb)] is plotted as a function of defect concentration at 300 K in Fig.~\ref{fig:kappa} (a). The $\kappa_\mathrm{L}$ calculated using ShengBTE is found to be in good agreement with existing literature.~\cite{Anand2019} The trend of lowering of $\kappa_\mathrm{L}$ with doping, as shown by both DC and BTE, is also found to be consistent with experiments.~\cite{Chai2015,Chauhan2019,Chauhan2019_1}

The cumulative $\kappa_\mathrm{L}$ is also plotted as a function of frequency of vibration of the phonons as shown in Fig.~\ref{fig:kappa} (b) for the parent and 12.5\% doped compositions of all the different compounds. It can be seen from the plot that the acoustic modes contribute to 90 \% of the $\kappa_\mathrm{L}$ in the case of the parents\textbf{~\cite{Tian2011}}. Since the $\kappa_\mathrm{L}$ reaches saturation at around the cut-off frequency of the acoustic modes [which is at $\sim$150 cm$^{-1}$ as shown in Fig.~\ref{fig:kappa} (b)]. Similar situation is observed in the case of ZrNi$_{1.125}$Sn and ZrPd$_{1.125}$Sn. In these cases, the acoustic modes are distinctly cut-off from the optical modes due to the avoided crossing created by the rattling mode of the isolated defect atom. Thus, in these cases, the saturation in $\kappa_\mathrm{L}$ is also reached at around the cut-off frequency ($\sim$50 cm$^{-1}$). On the other hand, in the case of ZrCo$_{1.125}$Sb, the extra Co atoms hybridize with the host lattice, as also discussed above. As a result of this, the energy carried by the acoustic modes is carried over a long-range of frequency [Fig.~\ref{fig:kappa} (b)] spreading to the optical modes. As a consequence of this, the cumulative $\kappa_\mathrm{L}$ is found to gradually increase up to a higher frequency (of $\sim$175 cm$^{-1}$) before it saturates. The scattering rates are plotted as a function of phonon frequencies in Fig.~\ref{fig:rate} (c) for the pristine and the 12.5 \% doped compositions. As expected, the scattering rates (which are inversely proportional to the scattering lifetimes) are found to be much higher for the doped compositions implying shorter scattering lifetimes. However, the scattering rates in ZrNi$_{1.125}$Sn and ZrPd$_{1.125}$Sn are found to be much higher as compared to ZrCo$_{1.125}$Sb. This is because of the difference in the defect chemistries in these compositions. In the case of the former two compounds, the introduction of the rattling modes of defect leads to softening of the phonon modes from $\sim$65 cm$^{-1}$ and $\sim$45 cm$^{-1}$ acoustic cut-off frequencies respectively (which varies as a function of the mass of the extra atom introduced). Because of this, the entire phonon spectrum (beyond the acoustic modes) comprises of a large number of localized flat modes. Thus, the scattering rates fluctuate drastically as a function of frequency in these compositions. In the case of ZrCo$_{1.125}$Sb, the scattering rates are found to be lower, which suggests the transfer of phonons to higher energy without getting scattered and hence implies a higher lattice thermal conductivity as compared to the other two compositions. The trend is again found to be in agreement with previous studies~\cite{Chauhan2019_1} as shown in Table~\ref{tab:kl}.

\subsection{Electronic band structure}
\begin{figure}
\centering
\includegraphics[scale=0.58]{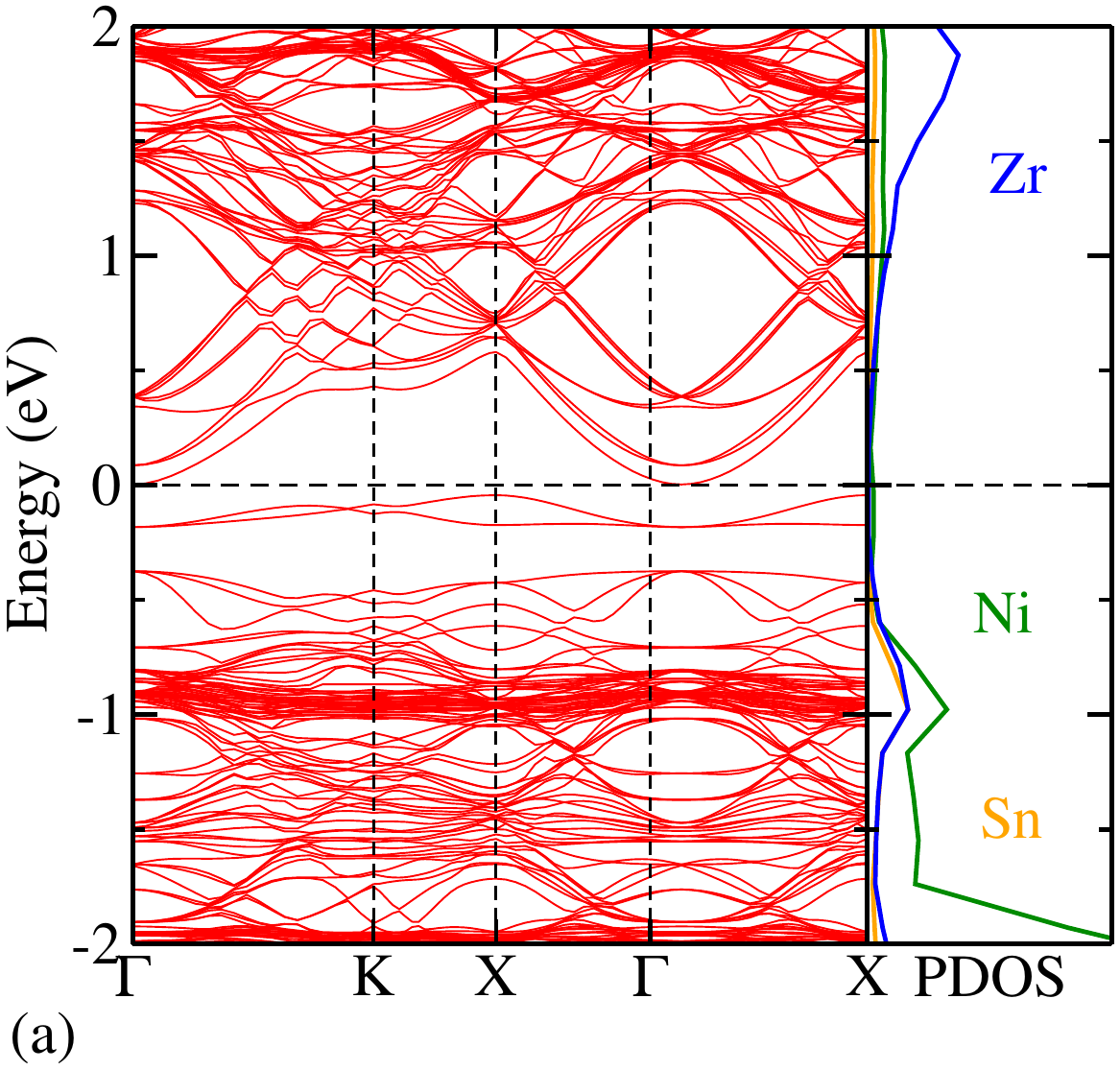}
\vspace{0.15cm}
\includegraphics[scale=0.58]{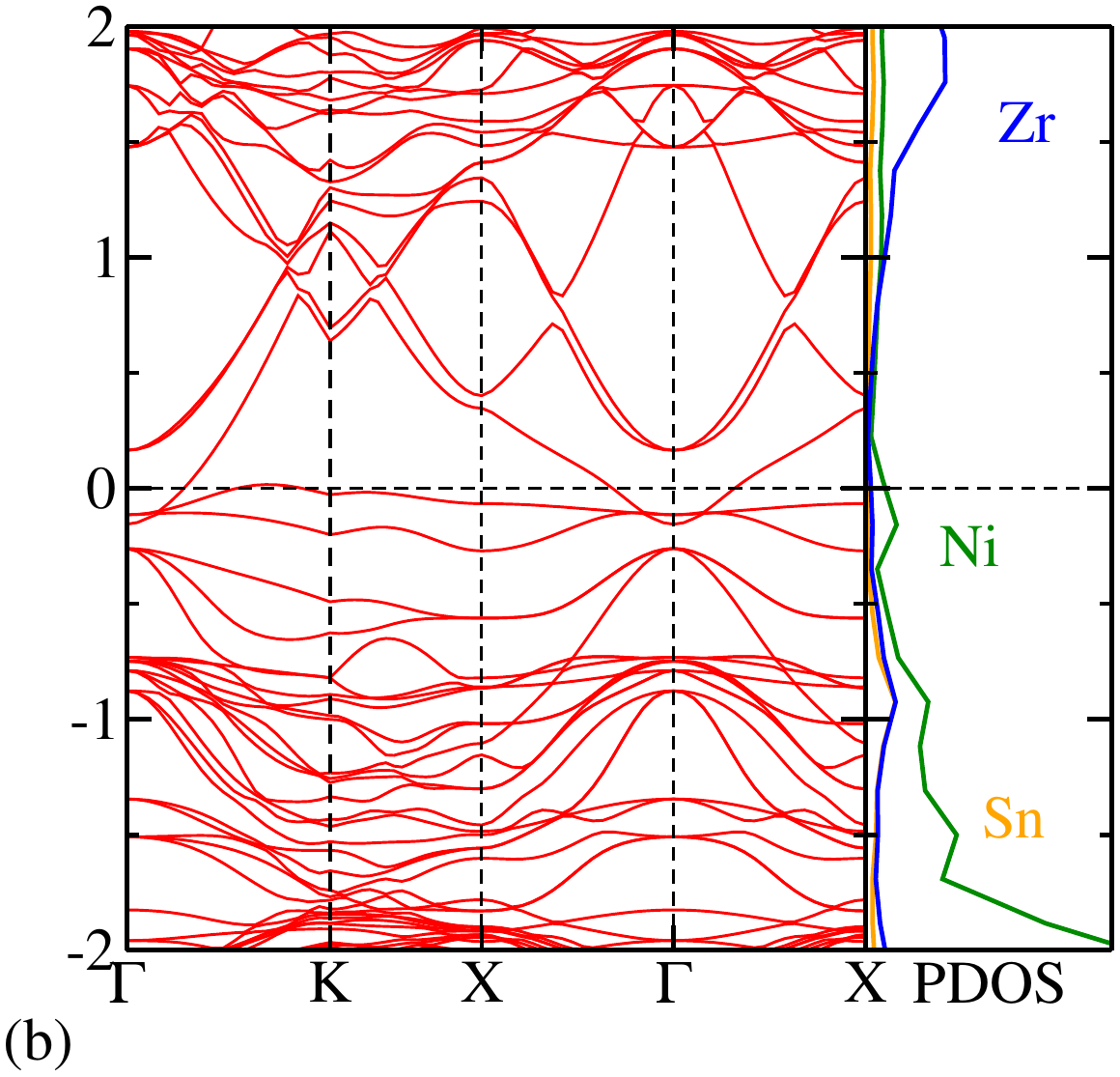}
\includegraphics[scale=0.58]{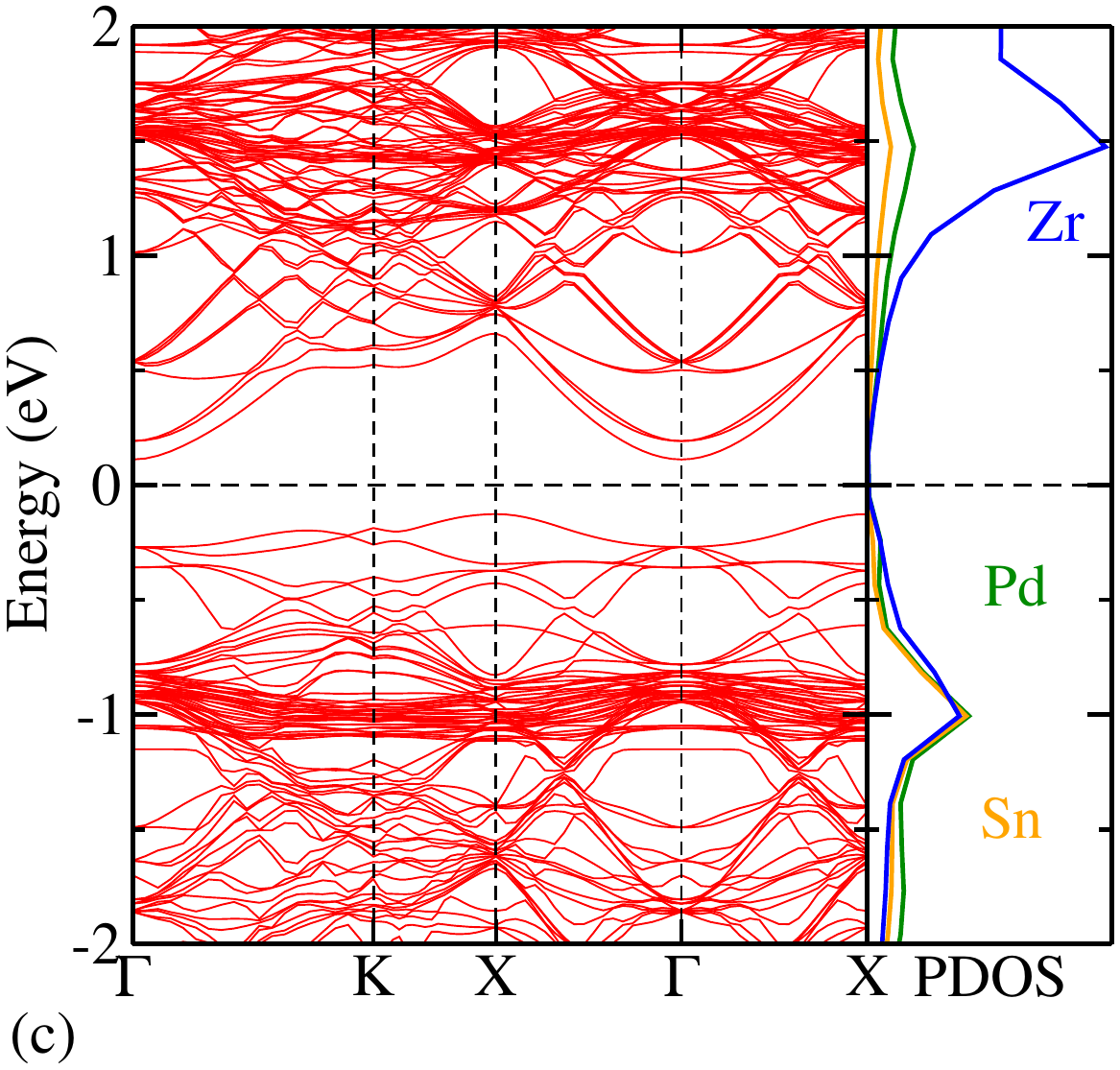}
\vspace{0.15cm}
\includegraphics[scale=0.58]{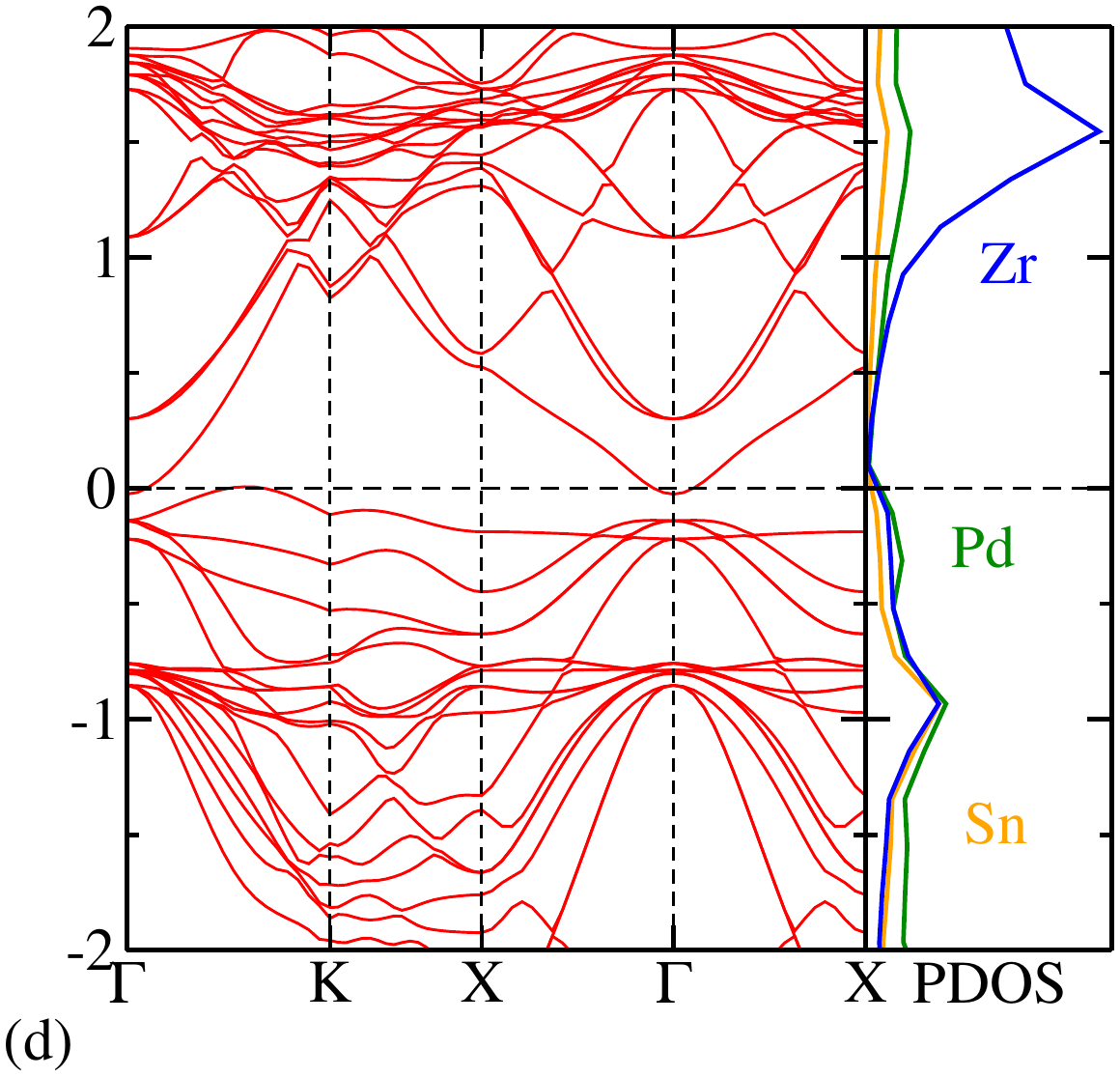}
\includegraphics[scale=0.58]{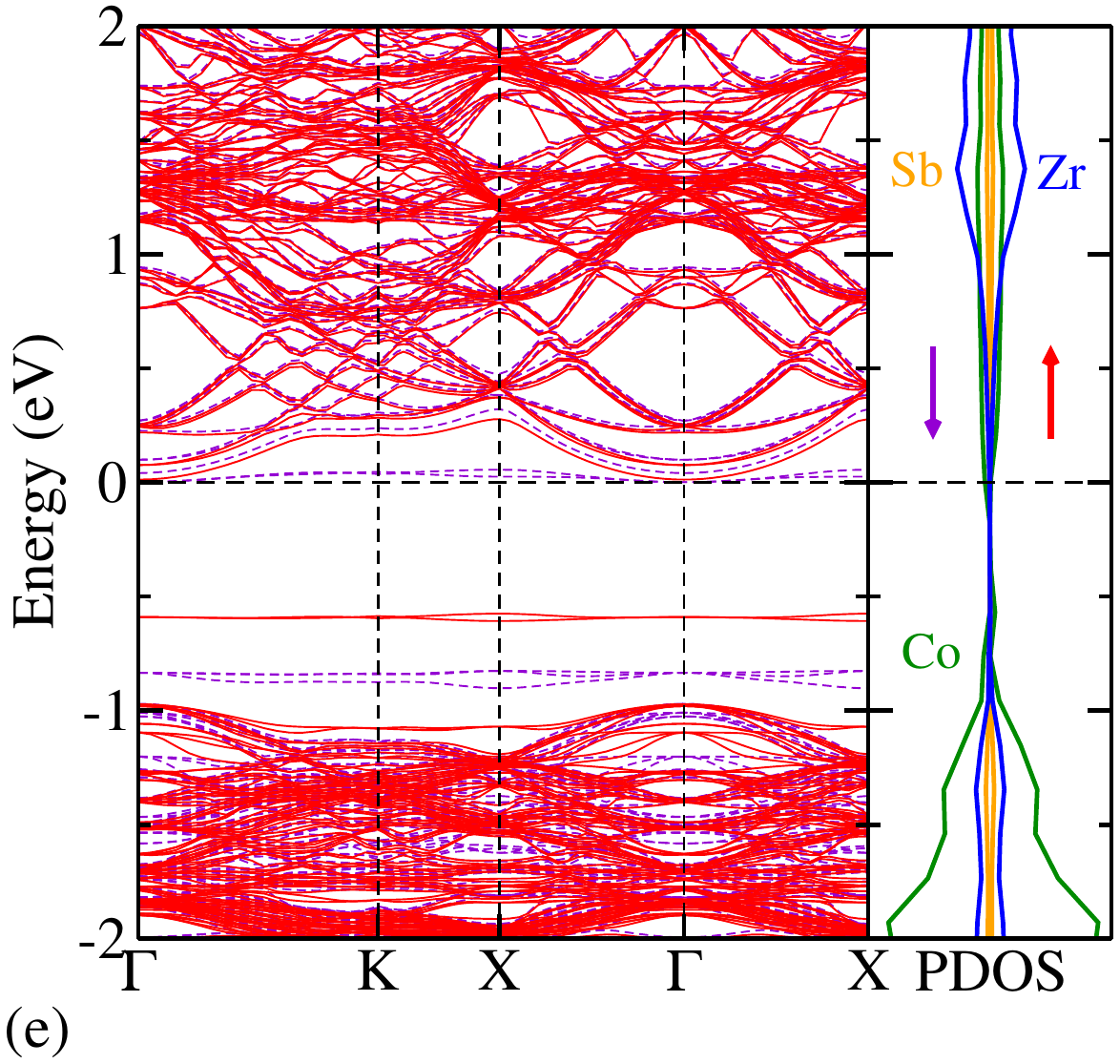}
\includegraphics[scale=0.58]{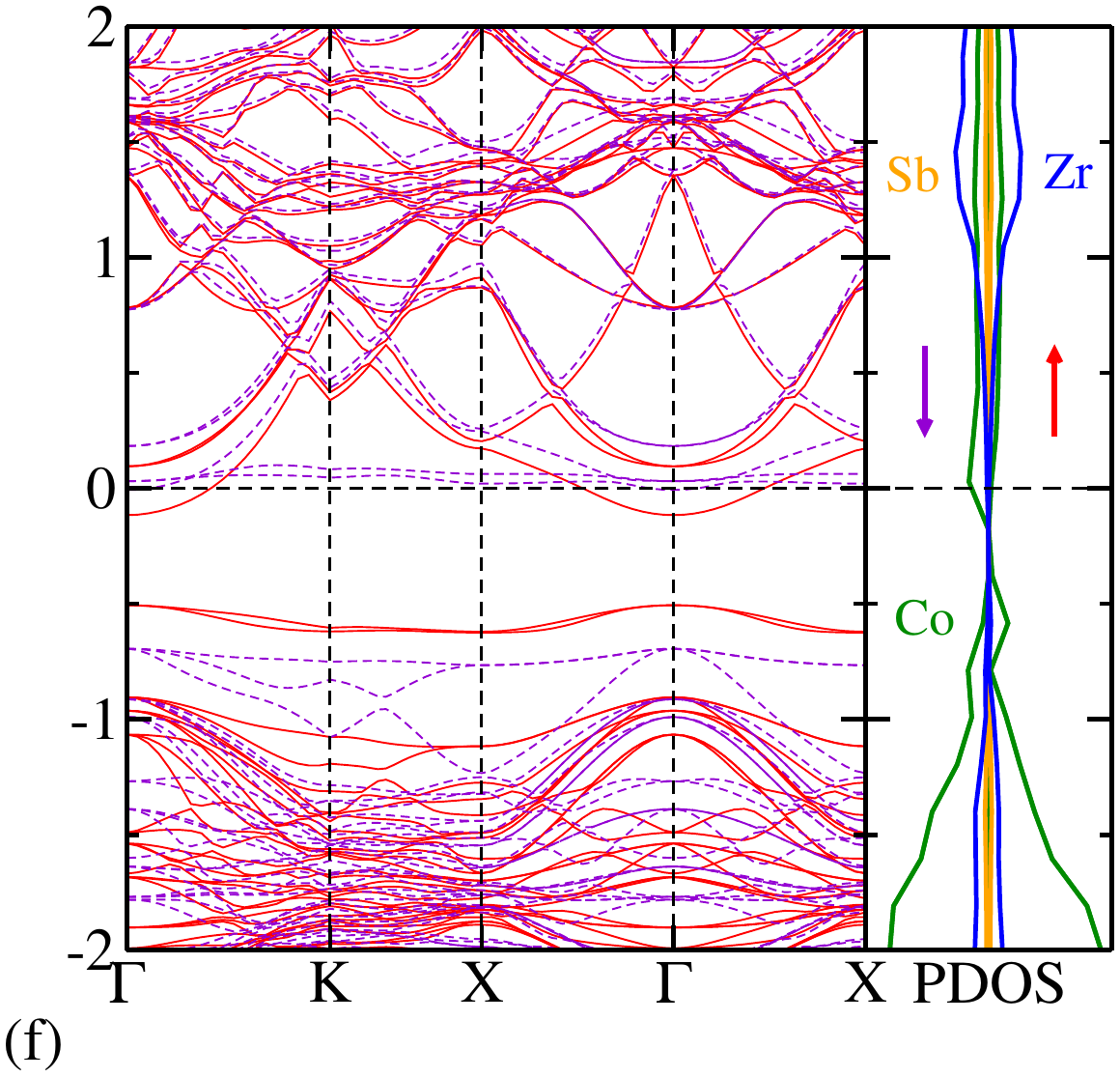}
\caption{Electronic band structure with partial density of states (PDOS) of (a) non magnetic ZrNi$_{1.03}$Sn [(b) ZrNi$_{1.125}$Sn] , (c) non magnetic ZrPd$_{1.03}$Sn [(d) ZrPd$_{1.125}$Sn], and (e) magnetic ZrCo$_{1.03}$Sb [(f) ZrCo$_{1.125}$Sb] (the partial-DOS of down and up spin components are shown separately). Fermi level is set to zero. (3 \% and 12.5 \% is formed using 2 $\times$ 2 $\times$ 2 supercell of conventional unit cell and primitive cell respectively, see section ``structure and stability'' for more details).}
\label{fig:bands}
\end{figure}

As can be seen from the vibrational analysis, all excess doping above 3 \% to 12.5 \% leads to the achievement of almost the lowest limit of $\kappa_\mathrm{L}$ for all the compositions. Thus, the conjecture is to analyze the tuning of the electronic transport coefficients while staying in this doping regime. For this purpose, the electronic band structure and the partial dos of the off stoichiometric compositions are investigated (see Fig.~\ref{fig:bands}). The pristine HH alloys under consideration in this work is found to be narrow bandgap semiconductors with bandgap ranging from $\sim$0.5-1.1 eV (see SI for Fig.~S4 in section S5). This bandgap is formed in between the five bonding (i.e., the t$_{2g}$ and e$_{g}$) and the five antibonding (i.e., the t$_{2g}^*$ and e$_{g}^*$) d-orbitals of the X and Y site atoms present in the compositions [see section S5 in SI]. The corresponding electronic band structure of the 3 \% [see Fig.~\ref{fig:bands} (a), (c), (e)] and 12.5 \% doped composition is shown in Fig.~\ref{fig:bands} (b), (d), and (f). Where few of the defect levels are formed inside the gap, while several other defect states are also found to be lying within the valence band (occupied deep levels). The identity, position and hybridization of the defect levels are discussed in detail in the following subsection.

Higher dispersion is observed in the midgap defect states (MIG) of 12.5 \% compositions [Fig.~\ref{fig:bands} (b), (d)] as compared to the 3 \% doping [Fig.~\ref{fig:bands} (a), (c)] in case of both the Ni as well as the Pd host. This is because of the increase in dopant-dopant interaction with the increase in their concentration in the lattice. The doping-induced changes in the Ni/Pd host are found to be similar. For the 3 \% doped Ni/Pd host, the MIGs are found to lie below the Fermi level resulting in the semiconducting [Fig.~\ref{fig:bands} (a), (c)] nature of the composition. On the other hand, in the case of 12.5 \% doped Ni/Pd host, the MIGs align in a way that the VBM as well as the CBm show partial overlap at the Fermi level. Thus, higher doping results in metallic behavior of the compositions. This doping-induced increase in metallicity in these compounds can also be related to the completely metallic behavior towards the FH phases (i.e., in ZrNi$_2$Sn and ZrPd$_2$Sn). In the case of Co doping, the defect states are found to be localized even for the 12.5 \% doped composition [Fig.~\ref{fig:bands} (f)]. Contrary to the other two doping cases, the defect states are found to be empty and therefore lie above the Fermi level. In the case of 12.5 \% composition, the MIGs move within the CB and a portion of the CB is found to cross the Fermi level leading to a light doping of the compound. The main difference, in this case arises from the wide band gap of the host i.e., ZrCoSb (of 1.1 eV), which results in the clear separation of the CB from the VB. Thus, out of all these doping scenarios, only 12.5 \% Co doing satisfies two conditions at once i.e., the light doping condition favorable for high power factor and low $\kappa_\mathrm{L}$. The bonding chemistry of the extra atom with the host lattice is discussed in details in the following section.

\subsection{Bonding analysis}
\label{sec:BA}
\begin{figure}[H]
\centering
\includegraphics[scale=0.45]{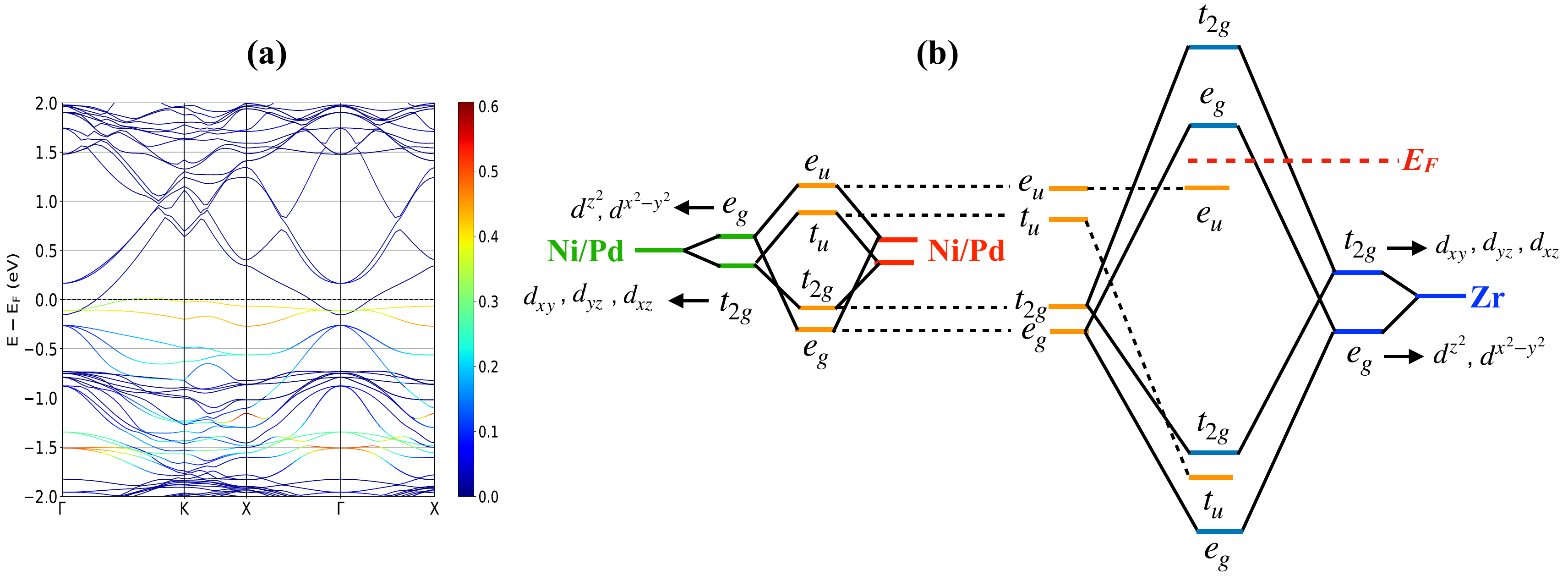} 
\includegraphics[scale=0.45]{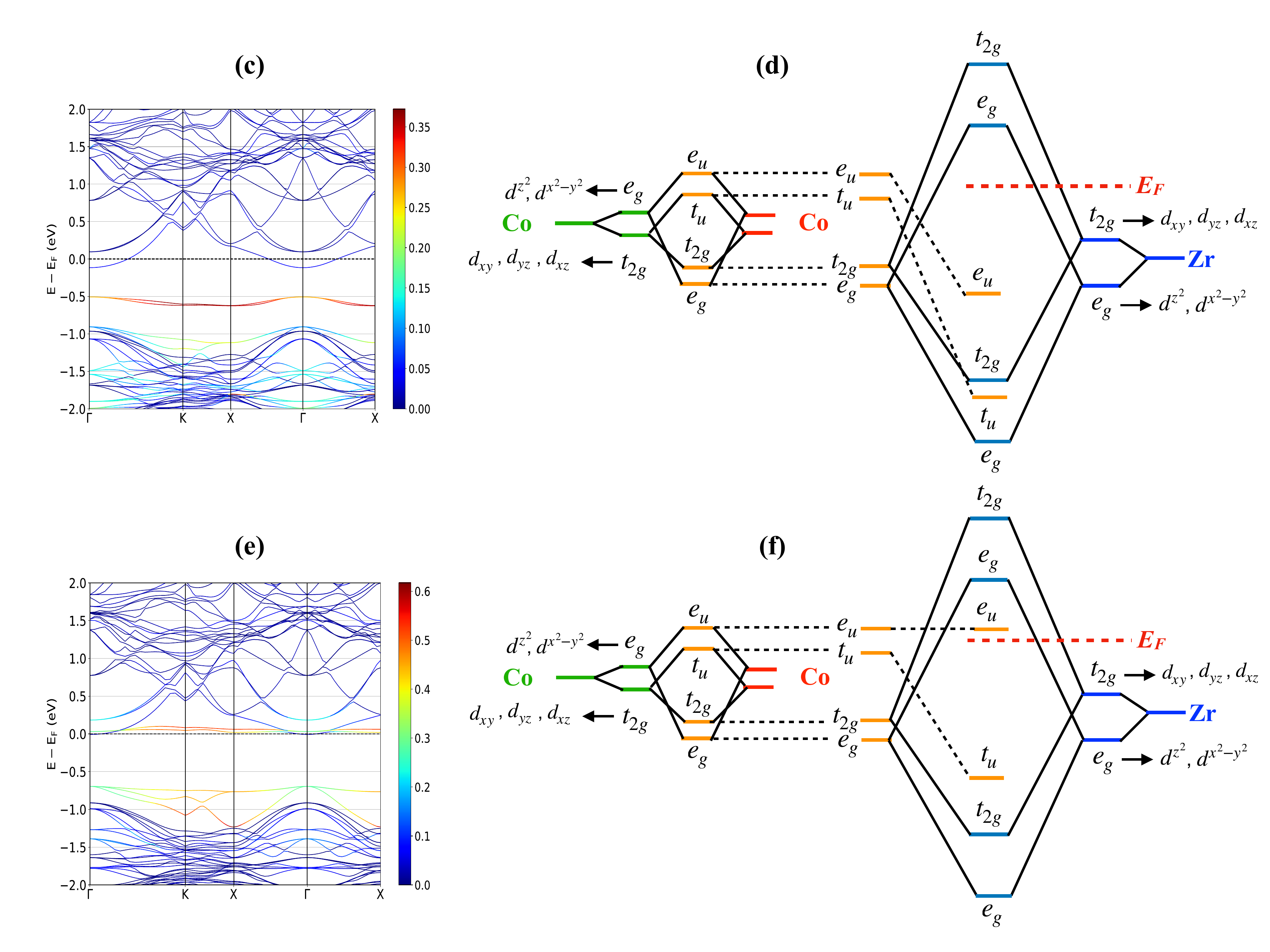} 
\caption{Electronic band structure (explicitly showing the orbital contribution stemming from $d$ orbitals of the extra atom highlighted using colors) and the corresponding schematic representation of the hybridization viz. (a-b) ZrNi$_{1.125}$Sn, (c-d) ZrCo$_{1.125}$Sb spin-up, and (e-f) ZrCo$_{1.125}$Sb spin-down. Here, e$_{g}$ (two-fold degenerate) denotes the d$_{z^{2}}$, d$_{{x^{2}}-{y^{2}}}$ orbitals and t$_{2g}$ refers to the (three-fold degenerate) d$_{xy}$, d$_{yz}$, d$_{zx}$ orbitals (see text for the detailed discussion).}
\label{fig:dstates}
\end{figure}

In half Hesuler alloys, only one atom is present at the Y-site and the entire crystal shows tetrahedral ($T_{d}$) symmetry. On the other hand, with addition of one more Y-site atom into the unit cell, the symmetry of the system changes. If one neglects the X and Z sites, the Y-site atoms have octahedral ($O_{h}$) symmetry with respect to each other [as shown in SI, Fig.~S5 (a) in section S5]. The addition of the extra atom in the host lattice changes the crystal symmetry and hence, the bonding chemistry.  Due to the extra atom, there is a significant charge rearrangement in the host HH matrix and this charge rearrangement occurs mainly at the nearby Y-site and X-site elements of the defect sites, which has $O_{h}$ and $T_{d}$ symmetry respectively w.r.t the extra dopant atom. An interesting perspective on the defect chemistry is given by Tolborg and Iversen using chemical bonding analysis, where they explained the bonding of the interstitial Ni in half Heusler by drawing analogy with Ni in full Heusler.~\cite{Tolborg2021} The extra atom forms a local full Heusler type environment, whereby they first hybridize with the Y site element and then with the X/Z site elements. Similar analogy is also used in the present study to explain the hybridization (as presented in Fig.~\ref{fig:dstates}) as discussed in details below.

Generally, in HH alloys, the $d$-states of the Y-site atom hybridize with the X-site ones in the unit cell and results in the formation of an energy bandgap between the five bonding states (three-fold degenerate, t$_{2g}$) and five anti-bonding states (two-fold degenerate, e$_{g}$) [as shown in SI, Fig.~S5 (b) in section S5] and the Fermi energy ($\mathrm{E}_{F}$) lies between these bonding and anti-bonding states. However, the inclusion of extra atoms results in the formation of ``in-gap'' states and shift the $\mathrm{E}_{F}$. The shifting in the $\mathrm{E}_{F}$ is found to be different for different hosts i.e. for non magentic Ni/Pd host is found to be different from the magnetic Co host.

The mechanism of orbital hybridization is explored in detail and shown schematically in Fig.~\ref{fig:dstates}. For the Ni and Pd-based compounds, each extra atom contributes ten electrons (i.e., $3d^{10}$ state electrons), which fill all the five $d$-states (i.e., the 3-t$_{2g}$ and 2-e$_{g}$). Due to the $O_{h}$ symmetry, as mentioned earlier, the triply degenerate t$_{2g}$ (d$_{xy}$, d$_{yz}$, d$_{zx}$) levels lie at a lower energy than the doubly degenerate e$_{g}$ (d$_{z^{2}}$, d$_{{x^{2}}-{y^{2}}}$) levels. Obeying the crystal symmetry, these orbitals hybridize with the orbitals of the nearest Ni/Pd atom, resulting in the formation of the bonding states (e$_{g}$, t$_{2g}$) and non-bonding states (e$_{u}$, t$_{u}$) as represented in Fig.~\ref{fig:dstates} (b). The possible energy diagram of the interactions between Ni-Ni (Pd-Pd) and Ni(Pd)-Zr atom is also illustrated in Fig.~\ref{fig:dstates} (b). The resulting bonding e$_{g}$ (t$_{2g}$) states of Ni hybridizes with the e$_{g}$ (t$_{2g}$) orbitals of Zr, thereby generating low energy bonding e$_{g}$ (t$_{2g}$) states and unoccupied anti-bonding states e$_{g}$ (t$_{2g}$). However, the energy levels of the non-bonding states of Ni remain unaffected as they do not hybridize with the Zr $d$-states. All these non-transforming states of Ni/Pd (e$_{u}$, t$_{u}$) are occupied and lie below the Fermi level, as shown in Fig.~\ref{fig:dstates} (b). The energy of t$_{u}$ is much lower than the e$_{u}$ states, therefore t$_{u}$ states are filled first and shift inside the valence band, at around $\sim$-1.5 for ZrNi$_{ 1.03}$Sn and $\sim$-2.7 for ZrPd$_{ 1.03}$Sn, whereas the e$_{u}$ bands are formed in the bandgap for Ni and closer to VBM for Pd, as clearly visible in the electronic band structure of ZrY$_{ 1+m}$Sn for $m$ = 0.03 and Y = Ni, Pd (see Fig.~\ref{fig:bands}). The major difference observed for the two cases is that in Zr-Ni-Sn systems, the e$_{u}$ and t$_{u}$ levels split and the e$_{u}$ levels are shifted to be formed in between the gap. However, in case of Zr-Pd-Sn systems the t$_{u}$ and e$_{u}$ levels are found to formed closed to the valence band (at about same energy), which is in confirmation with previous reports.~\cite{Do2014} However, with increase in dopant concentration (for $m$ = 0.125) the defect level interacts and becomes more dispersed as shown in Fig.~\ref{fig:dstates} (a).

In this case, spin polarized effect is observed for the small doping concentrations, due to the unpaired electrons. Each extra atom of Co, with electronic configuration $3d^{9}$, contributes nine $d$-state electrons to the system. The hybridization process between the orbitals of Co atoms is the same as discussed above for Ni/Pd atoms [see Fig.~\ref{fig:dstates} (d) or (f)]. Whereas, owing to the odd number of electrons, the non-transforming states e$_{u}$ of Co atom are not occupied for minority spins and therefore lie above the Fermi level and the t$_{u}$ states are completely occupied and lie below the Fermi level, near to the valence band maxima, as shown in Fig.~\ref{fig:dstates} (e). Moreover, for the majority spins e$_{u}$ states are occupied, lie near the VBM, whereas, the triply-degenerate t$_{u}$ states are found to lie deep inside the valence band, as shown in Fig.~\ref{fig:dstates} (c-d). In comparison, the scenario is a bit different for higher doping concentration (i.e.~of $m$ = 0.125). In higher doping compositions, a doping is also achieved resulting in the shifting of the $\mathrm{E}_{F}$ inside the CB. Thus, amongst all the different dopants and doping concentrations, only Co doping with 12.5 \% is found to result in a light doping of the system (i.e.~all the 3 \% doped compounds are found to intrinsic semiconductors and in all other 12.5 \% doped compounds, the MIGS are found to be completely dispersed from VBM to the CBm). This already implies towards a doping induced synergistic improvement in the electronic transport scenario.   

\subsection{Electronic transport properties}

\begin{figure}[H]
\centering
\includegraphics[scale=0.6]{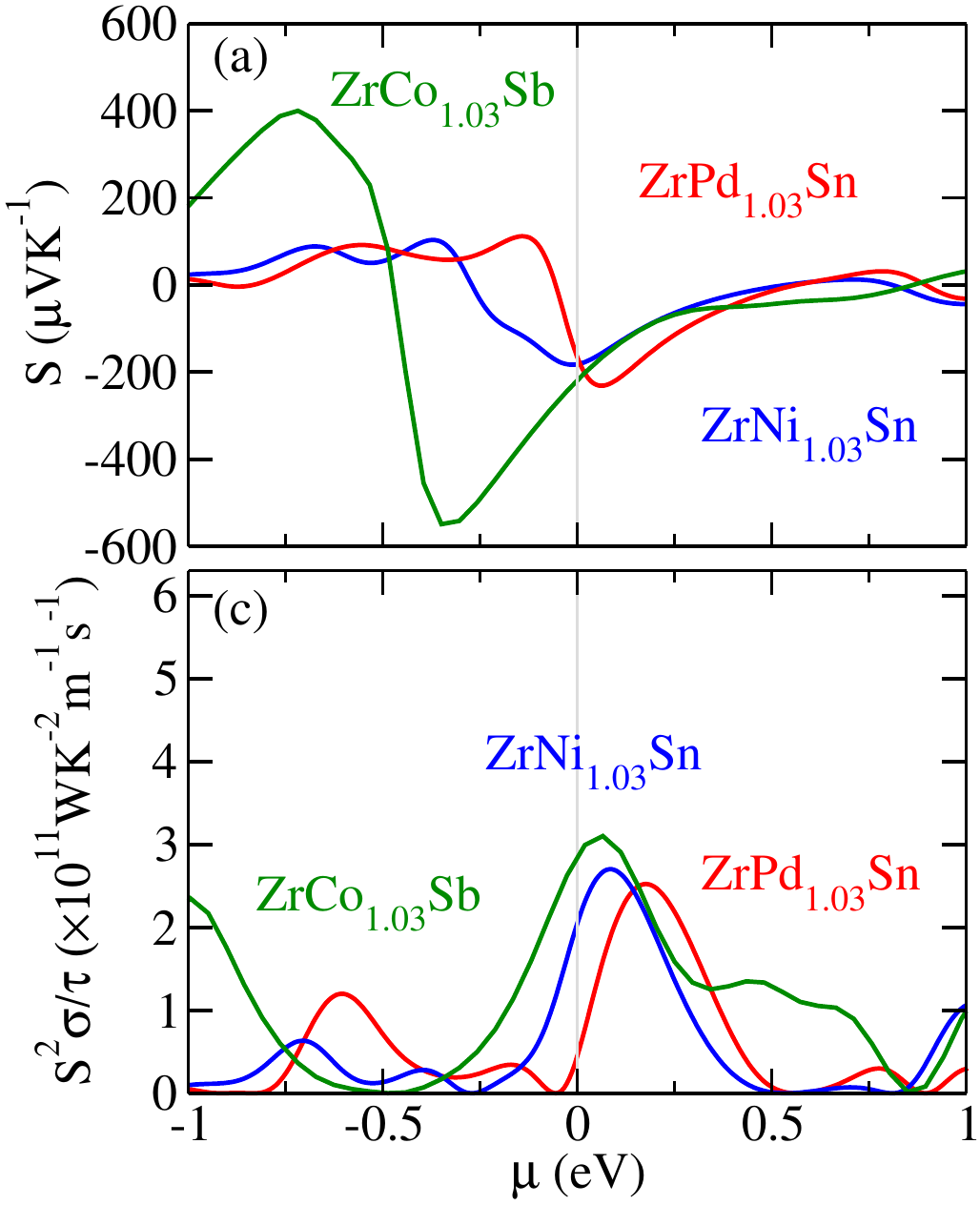}
\includegraphics[scale=0.6]{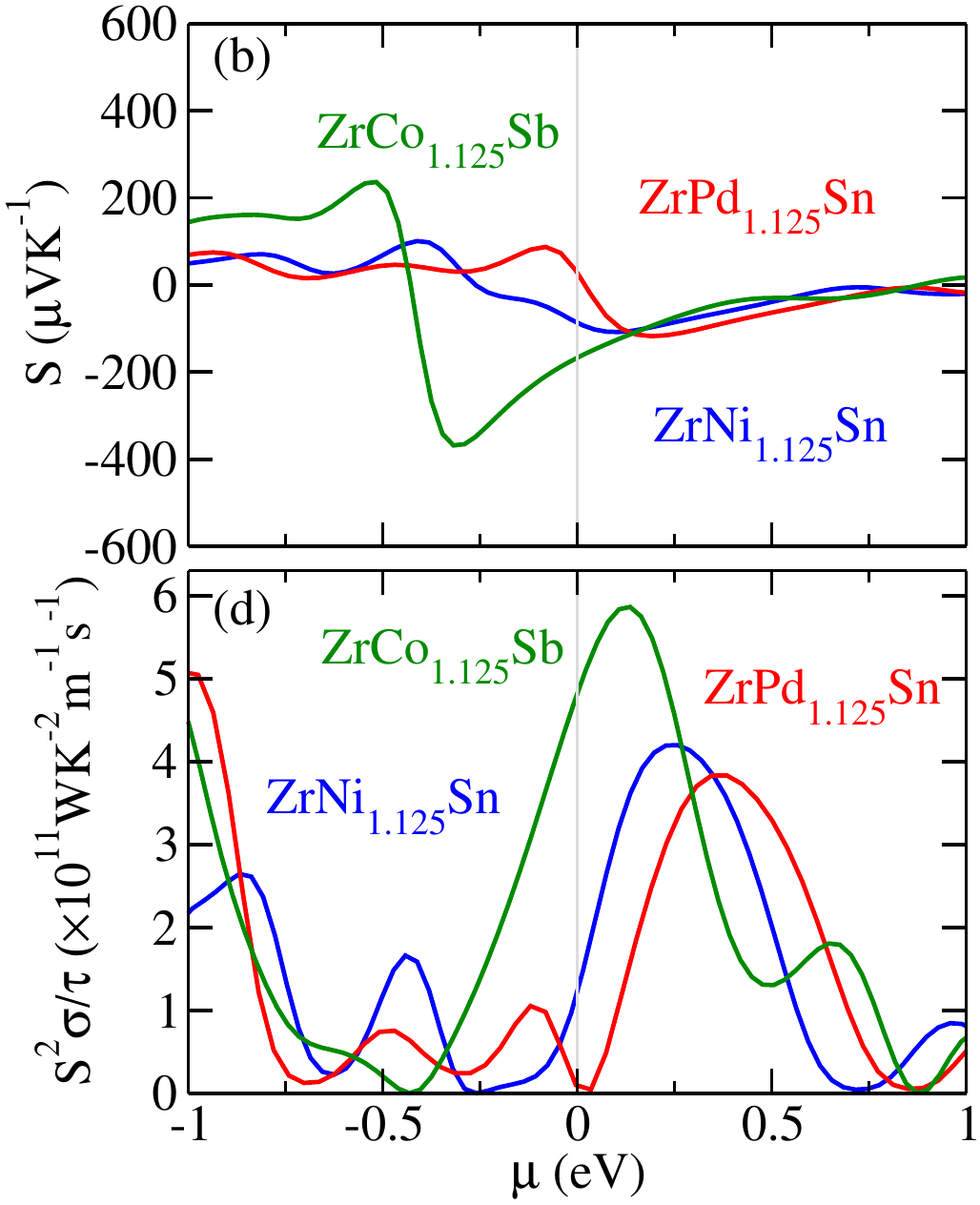}
\caption{(a-b) Seebeck coefficient ($S$) and (c-d) power factor ($S^{2}\sigma/\tau$) plotted as a function of electron chemical potential $\mu$ for the compositions viz. ZrY$_{1+m}$Z (Y = Ni, Pd, Co; Z = Sn, Sb) for $m$ = 0.03, 0.125 at 800 K.}
\label{fig:S}
\end{figure}

Finally, the transport properties of the 3 \% and 12 \% doped compositions are calculated using BTE for electrons. The corresponding Seebeck coefficient~($S$) and power factor, PF ($\frac{S^2\sigma}{\tau}$) scaled by the relaxation time are plotted as a function of chemical potential ($\mu$) of electrons and holes at 800 K (see Fig.~\ref{fig:S}). The corresponding $S$ and PF plots for the parent HHs are provided in the SI (see section S6). As seen from these plots, the parents are intrinsic semiconductors, which show $\frac{S^2\sigma}{\tau}|_{\mathrm{max}}$ at very high values of $\mu$ [viz.~$\sim$0.25 eV for ZrYSn (Y = Ni; Pd) while $\sim$0.7 eV for ZrCoSb] as shown in Fig.~S6 of section V in SI. In order to recover any meaningful PF, the host compounds have to be doped. However, for any meaningful doping that can be achieved experimentally (i.e., about 0.1 eV), the PF in the parent compounds is sufficiently low. Thus, the PF of the doped compositions is analyzed. All the doped compounds show n-type doping behavior, which is confirmed by the shift in the Fermi level towards the positive $\mu$ value. In Zr-Ni-Sn systems, the ZrNi$_{1.03}$Sn is found to show higher PF than the 12.5 \% compositions, which is in agreement with the experimental studies.~\cite{Chauhan2019, Chauhan2019_1}. However, as was expected from the electronic structure analysis of ZrCo$_{1.125}$Sb, a very high PF is attained compared to the 3 \% doped compositions following the perfect low doping situation prevailing in it.
However, these values of power factor is scaled by the electron scattering lifetime $\tau$. Explicit accurate estimation of $\tau$ from first principles is a very challenging task and has not been performed in this study, which may affect the actual power factor. A discussion of $\tau$, by comparing the scaled power factor with the experimental literature is given in the supplementary information (see section S6).

\section{Conclusions}
The effect of excess Y-site doping is explored on the vibrational and electronic transport properties of the ZrY$_{1+m}$Z ($m$ = 0, 0.03, 0.125; Y = Ni, Pd, Co; Z = Sn, Sb) compounds using first-principles based DFT calculations. A reverse conventional approach is followed to explore the efficient pathways to establish a synergy between the vibrational and electronic transport coefficients. For this purpose, the entire compositional space (for a varied doping concentration until the FH limit) is first explored to analyze the effect of doping on $\kappa_\mathrm{L}$. Once the optimal doping limit for achieving the low $\kappa_\mathrm{L}$ is identified, the electronic transport scenario in the regime is explored. The $\kappa_\mathrm{L}$ of the parent HHs is found to be drastically reduced (by at least 65 \%) as soon as doping of 3 \% by a Y site element is introduced in the vacant Y site positions of the HH lattice. For a higher concentration of doping to about 12.5 \%, the $\kappa_\mathrm{L}$ is found to reach the lowest. The lattice dynamics for these two doping concentrations is explored in detail, which suggests two different mechanisms to prevail for lowering the $\kappa_\mathrm{L}$. In the case of the isovalent Ni/Pd based compounds, the lowering in the $\kappa_\mathrm{L}$ is brought about by `rattling' modes of the extra dopant atom introduced, which do not couple with the host lattice. These modes act as scattering centers leading to disruption (due to avoided crossing) in the vibration of the acoustic phonons leading to the separation of the acoustic and the optical phonon modes. However, in the case of the Co-based host, hybridization between the extra dopant atom and the host lattice is observed which helps the acoustic modes in carrying the energy over a long-range in frequency. Naturally, the $\kappa_\mathrm{L}$ for the Co-hosts is found to be higher as compared to the other two hosts. However, all three cases of doping lead to lowering in the $\kappa_\mathrm{L}$. Subsequently, the effect of these dopings (viz.~3 \% and 12.5 \%) in the electronic structure and the electronic transport properties are also explored. For the 3 \% doping of the HH lattice, the compounds are found to retain their semiconducting properties, albeit some of the defect states of the extra atoms are found to be lying in the mid-gap (MIG). In case of higher doping concentration (of 12.5 \%), the Ni/Pd based compounds are found to show semiconductor to metal transition due to the overlap of the MIGs with the Fermi level. Only in the case of the Co-host (for the 12.5 \% doping), however, the MIGs are found to be completely unoccupied and a light doping is achieved by the crossing over the conduction band into the Fermi level. This may help in improving the transport scenario. Thus, this work provides a complete theoretical blueprint for achieving the highest efficiency for the entire compositional space via excess doping in laboratory. Needless to say, the present work may be extended to the entire half-Heusler class for improving their thermoelectric efficiency.\\

\begin{suppinfo}
S1: Formation energy analysis, S2: Optimized lattice parameter with doping concentration, S3: Phonon dispersion for 3 \% doping concentration, S4: Debye Callaway formalism, S5: Electronic band structure analysis and hybridization in pristine HH, S6: Electronic transport parameters of parents.
\end{suppinfo}


\begin{acknowledgement}
AB acknowledges the DST Inspire faculty project (DST/INSPIRE/04/2015/000089), IIT B seed grant project (RD/0517-IRCCSH0-043) and SERB ECRA project (ECR/2018/002356) for the financial assistance. The high performance computational facilities [viz. Aron (Ab-CMS lab, IITB), Dendrite (ME$\&$MS dept., IITB), and Spacetime, IITB] and CDAC Pune (Param Yuva-II) are acknowledged for providing the computational hours.
\end{acknowledgement}

\bibliography{achemso-demo}

\end{document}